\documentclass[12pt]{article}
\usepackage[utf8]{inputenc}
\usepackage{newtxtext} 
\usepackage{newtxmath} 
\usepackage{titlesec}

\titleformat{\section}
  {\normalfont}                 
  {\bfseries\thesection.}              
  {1em}                         
  {\bfseries}           

\titleformat{\subsection}
  {\normalfont}                 
  {\bfseries\thesubsection.}              
  {1em}                         
  {\bfseries\itshape}           

\titleformat{\subsubsection}
  {\normalfont}
  {\thesubsubsection.}
  {1em}
  {\itshape}

\usepackage[a4paper,top=2cm,bottom=2cm,left=3cm,right=3cm,marginparwidth=1.75cm]{geometry}

\usepackage[british]{babel}
\usepackage{microtype}
\usepackage[autostyle]{csquotes}
\usepackage{etoolbox}

\usepackage[
    backend=biber,
    style=ext-numeric-comp,
    sorting=none,
    maxnames=99,      
    giveninits=true,  
    doi=false,
    isbn=false,
    url=false,
    eprint=false
]{biblatex}

\DeclareFieldFormat[article]{title}{}

\DeclareFieldFormat[article]{year}{\mkbibparens{#1}}

\DeclareFieldFormat{journaltitle}{#1}
\DeclareFieldFormat{pages}{#1}
\DeclareFieldFormat{page}{#1}

\DeclareNameAlias{default}{given-family}

\DeclareBibliographyDriver{article}{%
  \printnames{author}%
  \setunit{\addcomma\space}%
  \printfield[journaltitle]{journaltitle}%
  \setunit{\addspace}%
  {\bfseries\printfield{volume}}
  \setunit{\addcomma\space}%
  \printfield{pages}%
  \setunit{\addspace}%
  \printfield{year}
  \finentry%
}

\DeclareBibliographyDriver{misc}{%
  \printnames{author}%
  \addcomma\space%
  preprint%
  \setunit{\addcomma\space}%
  \printfield{howpublished}%
  \space%
  \mkbibparens{\printfield{year}}\adddot%
  \newline
  $<$\printfield{url}$>$%
  \finentry
}

\DeclareFieldFormat{url}{\textnormal{#1}}

\DeclareBibliographyDriver{online}{%
  \usebibmacro{bibindex}%
  \usebibmacro{begentry}%
  \printnames{author}%
  \addcomma\space
  \printfield{title}%
  \space\mkbibparens{\printfield{year}}%
  \adddot\newline
  \printfield{url}%
  \space
  \mkbibparens{accessed \printfield{month}\addspace\printfield{urlday}\addcomma\addspace\printfield{urlyear}}%
  \adddot
  \usebibmacro{finentry}%
}

\DeclareNameAlias{default}{given-family}

\DeclareFieldFormat[book]{title}{\mkbibemph{#1}}

\DeclareFieldFormat[book]{pages}{pp.\space #1}

\DeclareBibliographyDriver{book}{%
  \printnames{author}%
  \setunit{\addcomma\space}%
  \printfield{title}%
  \setunit{\addspace}%
  \printtext{%
    \mkbibparens{%
      \printlist{publisher}%
      \addcomma\space%
      \printlist{location}%
      \addcomma\space%
      \printfield{year}%
    }%
  }%
  \iffieldundef{pages}
    {}
    {%
      \setunit{\addcomma\space}%
      \printfield{pages}%
    }%
  \finentry
}

\usepackage{array}

\usepackage[hidelinks]{hyperref}
\usepackage{amsmath}
\usepackage{bm}
\usepackage{url}
\usepackage{mathtools}
\usepackage{float} 
\usepackage{pdfpages}
\usepackage{authblk}

\AtBeginBibliography{\sloppy}

\newcommand{\op}[1]{\hat{\mathrm{#1}}}

\newcommand{\crt}[1]{\hat{a}_{{#1}}^{\dag}}
\newcommand{\anh}[1]{\hat{a}_{{#1}}}
\newcommand{\crtt}[1]{\tilde{\hat{a}}_{{#1}}^{\dag}}
\newcommand{\anht}[1]{\tilde{\hat{a}}_{{#1}}}

\newcommand{\ket}[1]{|{#1}\rangle}
\newcommand{\braket}[2]{\langle {#1}| {#2}\rangle}
\newcommand{\mel}[3]{\langle {#1}| {#2}| {#3} \rangle}
\DeclarePairedDelimiterX{\normman}[1]{\lVert}{\rVert}{#1}

\title{Trotter Error and Orbital Transformations in Quantum Phase Estimation}

\author[1]{Marvin Kronenberger}
\author[1]{Mihael Erakovic}
\author[1,\,*]{Markus Reiher}

\affil[1]{ETH Zürich, Department of Chemistry and Applied Biosciences, Vladimir-Prelog-Weg 2, 8093 Zürich, Switzerland}
\affil[*]{Corresponding author: \href{mailto:mreiher@ethz.ch}{mreiher@ethz.ch}}

\date{February 20, 2026}

\begin{document}
\maketitle

\begin{abstract}
Quantum computation with Trotter product formulae is straightforward and requires little overhead in terms of logical qubits. The choice of the orbital basis significantly affects circuit depth, with localised orbitals yielding lowest circuit depths. However, literature results point to large Trotter errors incurred by localised orbitals. Here, we therefore investigate the effect of orbital transformations on Trotter error. We consider three strategies to reduce Trotter error by orbital transformation: (i) The a priori selection of an orbital basis that produces low Trotter error. (ii) The derivation of an orbital basis that produces a ground state energy free of Trotter error (as we observed that the Trotter error is a continuous function in the Givens-rotation parameter, from which continuity of this error upon orbital transformation can be deduced). (iii) Application of propagators that change the computational basis between Trotter steps. Our numerical results show that reliably reducing Trotter error by orbital transformations is challenging. General recipes to produce low Trotter errors cannot be easily derived, despite analytical expressions which suggest ways to decrease Trotter error. Importantly, we found that localised orbital bases do not produce large Trotter errors in molecular calculations, which is an important result for efficient QPE set-ups.
\end{abstract}

\section*{Keywords}
quantum computing; quantum chemistry; Trotter error; orbital basis; randomised product formula

\section{Introduction}\label{sec:intro}
Accurate quantum chemical calculations of highly correlated electronic structures in chemistry
require high-accuracy multi-configurational methods \cite{baiardi2020density, evangelista2018perspective, sokolov2024multireference, park2020multireference}.
The exponential scaling of the Hilbert space dimension with the size
of the active orbital space remains a key hurdle for the development of such methods on
classical hardware.
Quantum computers have an exponential advantage in the number
of quantum versus classical bits required to encode elements
of this Hilbert space \cite{QuantCompBible}.
This motivates the development of algorithms on quantum hardware
which allow for tackling strongly correlated
molecules.

Quantum phase estimation (QPE) \cite{Kitaev1995} estimates the eigenenergies of arbitrary
Hamiltonians on fault-tolerant quantum hardware \cite{AbramsLloyd} and
requires the implementation of a unitary propagator $U$,
whose eigenphases $\omega_k$ deliver the eigenenergies $E_k$ of
the Hamiltonian $\op{H}$ \cite{AbramsLloyd, AGuzik2005}.
The probability with which a binary representation of
each eigenphase $\omega_k$ can be measured
is $\lvert \braket{\psi(0)}{\psi_k} \rvert^2$ \cite{Whitfield2011}.
Here, $\ket{\psi(0)}$ denotes the input state which will be evolved
in time via Hamiltonian simulation and $\ket{\psi_k}$ is
the eigenstate of $\op{H}$ corresponding to the eigenenergy $E_k$.

To compute the ground state energy $E_0$ to some accuracy $\epsilon$,
two quantities determine the total quantum-resource cost 
of the QPE circuit.
First, the number of qubits and quantum gates required
to prepare an initial state $\ket{\psi(0)}$,
which has sufficiently high overlap with the target ground state wave function $\ket{\psi_0}$ (state-preparation cost).
Second, the number of qubits and quantum gates required
to implement the unitary propagator $U$  
(Hamiltonian-simulation cost).

Within QPE, Hamiltonian simulation for time $t_{\mathrm{total}}$
determines the ground state energy $E_0$ to a desired
accuracy $\epsilon$ (where $t_{\mathrm{total}} \propto 1/\epsilon$)
by implementation of the unitary
operation $U=\exp (-i\op{H}t_{\mathrm{total}})$ on a quantum device.
A plethora of choices for the numerical representation of
the Hamiltonian $\op{H}$ exists \cite{HelgakerBible, Reiher2017, MartinezPartitionings, QubitizationSingleFact, Reiher2021, THC2019, MottaLowRank, peng2017decomp, babbush_large_first_quant_paper, first_quant_any_basis, rocca2024reducing, low2025fast}.
This choice directly affects
both the cost of state preparation \cite{erakovic2025high, morchen2024classification, tubman2018postponing, berry2018improved, huggins2025efficient} and the Hamiltonian
simulation itself \cite{first_quant_any_basis, babbush_large_first_quant_paper, Reiher2021, QubitizationSingleFact, THC2019, low2025fast, MartinezPartitionings, rocca2024reducing}.
Here, we consider the second-quantised electronic Hamiltonian expressed in a spin orbital basis,
which is commonly employed in the literature \cite{AGuzik2005, Whitfield2011, Reiher2017, mehendale2025estimating, THC2019}.

Various schemes are available to implement the
time-evolution operator $U$ on a quantum computer.
A Trotter product formula can be employed to split
the exponential of the operator sum into products
of single operator exponentials \cite{Trotter1959, suzuki1976generalized, AbramsLloyd, AGuzik2005}
\begin{equation}\label{TrotterFormula_1}
U=\exp (-i\op{H}t_{\mathrm{total}}) = \exp \left( {\sum_k^{\Gamma} -i\op{H}_k t_{\mathrm{total}}} \right)
=\lim_{s \to \infty} \prod_k^{\Gamma} \left( \exp \left( -i\op{H}_k t \right)\right)^s \ ,
\end{equation}
where $t = {t_{\mathrm{total}}}/{s}$
denotes the time evolution $t$ associated with each so-called \enquote{Trotter step}.
The single-operator exponentials $\exp(-i \op{H}_k t)$ can then be translated
into quantum circuits as described, for example, in Refs. \cite{Whitfield2011, seeley2012bravyi}.
In practice, the number of Trotter steps $s$ (\enquote{Trotter step number})
in Eq. \eqref{TrotterFormula_1}
should be chosen such that the error $\lvert \Delta E_0 \rvert$ of the
ground state energy $E_0$ is smaller than the target accuracy.
For electronic structure calculations, this is typically chemical accuracy, which
is characterised by Trotter errors below $0.1-1$ $\mathrm{m}E_{\mathrm{h}}$ \cite{Reiher2017, BabbushOrbInfluence}.

Alongside trotterisation, several other strategies exist to
propagate the wave function in the context of QPE.
Implementing Linear Combinations of Unitaries (LCU)
on a quantum circuit \cite{LCUOriginal2012}
allows for the representation of the time evolution operator via
multi-product formulas (MPFs) \cite{LCUOriginal2012, MPFBravyi2024}.
MPFs refer to linear combinations of trotterised propagators with different
Trotter steps $s$. Since the sums of unitary operators are not guaranteed
to construct a unitary operator, MPFs are generally non-unitary
propagators and require additional ancilla
qubit overhead \cite{LCUOriginal2012}.
Hamiltonian simulation is also possible with
\enquote{qubitised} block encodings of the Hamiltonian \cite{LowQubitization}.
Evolving the wave function with the \enquote{walker} operator
$\mathcal{W}[\op{H}]=\exp (-{i\arccos{\op{H}/\lambda}})$ \cite{Reiher2021, babbush2018encoding, berry2018improved, poulin2018quantum}
instead of $U=\exp (-i\op{H}t)$ allows error-free calculations of $E_0$ from QPE.
The 1-norm $\lambda$ denotes a normalisation constant, such that the spectrum
of $\op{H}/\lambda$ is normalised.
While benchmark studies \cite{Reiher2021, THC2019, rocca2024reducing, ChristandlRandom2025, low2025fast}
on proposed target molecular systems \cite{Reiher2017, GChanFeMoCo2019, Reiher2021, goings2022reliably, hariharan2024modeling, genin2022estimating, morchen2024classification, lee2023evaluating, maskara2025programmable}
show lower Toffoli gate counts compared to trotterised methods,
this comes at greater than approximately an order 
of magnitude increase
in logical qubit requirements.
The additional qubit requirements of qubitisation stem from
the necessity to block-encode $\op{H}/{\lambda}$, which requires the embedding of
the Hamiltonian in a larger Hilbert space \cite{LowQubitization}.

The smaller number of logical qubits required for trotterisation
makes Hamiltonian simulation with trotterised product formula
well suited for 
early future fault-tolerant hardware, which will be qubit limited.
For this reason, we focus here on strategies to reduce the cost
of trotterised Hamiltonian simulation.
As is apparent from Eq. \eqref{TrotterFormula_1},
lower Trotter step numbers $s$ and a lower number of
terms $\Gamma$ in the Hamiltonian
decrease the resource cost of trotterised Hamiltonian
simulation.
To allow for the use of larger Trotter step numbers $s$ necessitates
low Trotter errors $\lvert \Delta E_0 \rvert$ in the ground state energy $E_0$.
Higher-order Trotter formulae
reduce the Trotter error, but come at an exponential increase
in the circuit depth of a Trotter step \cite{ChildsTrotter, childs2019nearly}.
First- and second-order Trotter formulae have therefore been
predominantly employed in the literature \cite{Reiher2017, BabbushOrbInfluence, mehendale2025estimating, poulin2015trotter}.

A variety of factors have been investigated
to reduce Trotter error (independent of the order of the employed Trotter formula):
Specifically, the choice of orbital basis \cite{BabbushOrbInfluence},
ordering of terms in the Trotter series \cite{TrotterOrderingImpact}
(which we denote \enquote{(Trotter) series-ordering}),
and Hamiltonian representation \cite{MartinezPartitionings}
on Trotter error have been analysed.
In recent years, several approaches for Hamiltonian simulation
with product formula beyond fixed-order Trotter formula
have been suggested.
Ref. \cite{luo2025efficient} applies tensor
hypercontraction \cite{THC_ref1, THC_ref2, THC_ref3, THC2019}
to construct the smallest possible basis, in which the electron repulsion
integral tensor is diagonal.
This facilitates error-free simulation of the two-body
interactions in the Hamiltonian at the cost of
additional qubits due to the increased basis size.

State-of-the-art simulations of electronic
Hamiltonians by product formulae exploit partially
randomised product formulae \cite{JinPartRandom, hagan2023composite, ChristandlRandom2025}.
The idea to approximate Hamiltonian simulation with
randomised product formulae instead of fixed-order Trotter decompositions
followed from work by Childs et al. \cite{ChildsRandomization}
and by Campbell \cite{qDRIFT}.
The former authors derived tighter bounds on Trotter error
for propagators with random reorderings of the terms $\op{H}_k$ compared to fixed
series order propagators \cite{ChildsRandomization}.
The quantum stochastic drift (qDRIFT) protocol \cite{qDRIFT}
abandons Trotter series altogether and 
instead approximates Hamiltonian simulation
by a Markovian process, where the operators
$\exp{\left(-i \op{H}_k t \right)}$
are randomly sampled according to the weights $\lvert h_k \rvert$
in $\op{H}_k$ of Eq. \eqref{Ham2nd}.
These weights $\lvert h_k \rvert$ depend on the specific
representation of the Hamiltonian terms $\op{H}_k$ that are employed.
They generally correspond to weighted sums of the
integrals $h_{ij}$ and $h_{pqrs}$, as, for instance, detailed
in Ref. \cite{OrbMin1norm}.
Hamiltonian simulation with partially randomised product
formula implement terms with large weights
by high-order Trotter formulae.
Terms with small weights are evolved
by the qDRIFT protocol \cite{ChristandlRandom2025}.
The cost of Hamiltonian simulation with qDRIFT scales
linearly with the squared Hamiltonian 1-norm $\lambda^2$,
where $\lambda = \sum_k \lvert h_k \rvert$.
It is well established that localised orbital bases minimise
the Hamiltonian 1-norm \cite{OrbMin1norm}
and therefore achieve the lowest-cost Hamiltonian
simulation with qDRIFT.

Finding an optimal orbital basis in which to implement the large coefficient terms is, however, less straightforward.
Ref. \cite{BabbushOrbInfluence} is the only
study which thoroughly presents numerical data of Trotter error as a function
of commonly employed orbital bases.
This study \cite{BabbushOrbInfluence} reported comparatively large Trotter errors in local orbital bases, 
which is in contrast with the
known correlation between upper
bounds of Trotter error and sums of the absolute values 
of the
Hamiltonian weights, i.e., the 1-norm $\lambda$ \cite{Reiher2017, ChildsTrotter, ChristandlRandom2025}.
Since the difference between canonical and localised orbitals
is more pronounced for extended molecular systems \cite{LocalityQuantChem, OrbMin1norm},
we provide additional results on
the effect of orbital localisation on Trotter error
as part of this work.
We investigate Trotter errors for active spaces of linear and non-linear
$\pi$-conjugated molecules, $\pi$-systems in short, where the difference between canonical (i.e., delocalised) and
localised orbitals is more pronounced compared to atomic,
diatomic, and triatomic systems
investigated in Ref. \cite{BabbushOrbInfluence}.

Following Childs et al. \cite{ChildsRandomization}
and Mart{\'\i}nez-Mart{\'\i}nez et al. \cite{MartinezPartitionings},
we also study randomised propagators.
We analyse whether randomly changing the
series-ordering or the orbital basis in each Trotter step
reduces the Trotter error in the ground state energy.
We emphasise that achieving error reduction by a dynamic
orbital basis (compared to dynamic Trotter series-ordering) can be
advantageous, since it allows for the selection of a fixed Trotter series order.
To reduce circuit depth, a Trotter series order which maximises
the number of gate cancellations could then be
employed \cite{TrotterLexicographicOrder, mukhopadhyay2023synthesizing}.
We employ a Givens rotation representation of
the orbital basis to account for a basis change.
Examples where error cancellation, averaging effects,
and error magnification can be observed are discussed.
Finally, we discuss whether the continuity of Trotter error
upon orbital transformations can be leveraged to determine
an orbital basis with little or no Trotter error in the ground
state energy.

\section{Theory}\label{sec:theory}

In this section, we introduce the
representations of the electronic Hamiltonian employed
throughout this work.
We describe how orbital transformations can be leveraged
to change the basis in which the electronic Hamiltonian is expressed.
We provide an overview on methods to assess Trotter error.
We discuss which of these methods are suited
to compare Trotter errors between different orbital bases
for molecular systems and what easily obtainable quantities,
which we refer to as descriptors hereafter,
might correlate with Trotter error.
The identification of such descriptors is desirable,
since it would enable the selection of a low-Trotter-error
orbital basis prior to an actual quantum computation.
Finally, Section \ref{subsec:randomized_propagator_deriv} introduces the
randomised propagators studied in this work.

\subsection{Representations of the electronic Hamiltonian}\label{subsec:ham_rep}

Throughout this work, we employ second-quantised electronic
Hamiltonians, which are defined in Eq. \eqref{Ham2nd} for a system of $M$ electrons and
$K$ nuclei expressed in a basis of $N$ spin orbitals \cite{HelgakerBible}:
\begin{equation}\label{Ham2nd}
\op{H} = \sum_{i,j}^N h_{ij} \crt{i}\anh{j}
+ \frac{1}{2} \sum_{p,q,r,s}^N h_{pqrs} \crt{p}\crt{q}\anh{r}\anh{s}
\equiv
\sum_k^{\Gamma} \op{H}_k \ .
\end{equation}
$\crt{i}$ and $\anh{i}$ denote the fermionic creation and
annihilation operators, respectively.
The one- and two-electron integrals $h_{ij}$ and $h_{pqrs}$
are defined as in Ref. \cite{Whitfield2011} in the below
equations:
\begin{equation}\label{one_body_integral}
h_{ij} = \int  \chi_i^* (\mathbf{x})
\left( -\frac{1}{2} \nabla^2 - \sum_A^K \frac{Z_A}{r_{\mathbf{x}, A}} \right)
\chi_j (\mathbf{x}) \ {\mathrm{d} \mathbf{x}},
\end{equation}
\begin{equation}\label{two_body_integral}
h_{pqrs} = \int 
\frac{\chi_p^* (\mathbf{x_1}) \chi_q^* (\mathbf{x_2})
\chi_r (\mathbf{x_2})  \chi_s (\mathbf{x_1})}{r_{1,2}}
\ {\mathrm{d} \mathbf{x_1}} {\mathrm{d} \mathbf{x_2}} \ .
\end{equation}
The set of functions $\{\chi_i(\mathbf{x})\}_{i=1}^{N}$
denote the spin orbitals, $r_{\mathbf{x}, A}$ denotes
the distance between the electron and the $A$'th nucleus,
and $r_{12}$ the distance between two electrons.
The charge of nucleus $A$ is defined as $Z_A$, and
$\mathbf{x} = (\mathbf{r}, \sigma)$ is a vector of the spatial coordinate $\mathbf{r}$
and spin $\sigma$ of an electron.
The spin orbitals are obtained
as a product of orthonormal spatial
orbital functions $\{\Phi_{i}(\mathbf{r})\}_{i=1}^{N/2}$, spatial orbitals in short,
and a spin-function $s(\sigma)$ which assigns either $\alpha$ or $\beta$
spin to a spin orbital
\begin{equation}\label{spinorb_defintion}
\chi_i(\mathbf{x}) =  \Phi_{i} (\mathbf{r}) s_i(\sigma).
\end{equation}
The spatial orbitals are obtained from the linear combination of atom-centered spatial functions
$\{\phi_{\mu}(\mathbf{r})\}_{\mu=1}^{N/2}$,
\begin{equation}\label{SPATORB}
\Phi_{i} = \left( \sum_{\mu=1}^{N/2} C_{i, \mu} \phi_{\mu}(\mathbf{r}) \right) \ ,
\end{equation}
where the elements $C_{i, \mu}$ of the coefficient matrix $\mathbf{C}$ are the spatial orbital coefficients and
store the contribution of each atom-centered spatial function $\phi_{\mu}(\mathbf{r})$
to the spatial orbital $\Phi_i$.

Throughout this work, two representations of the
Hamiltonian terms $\op{H}_k$ in Eq. \eqref{Ham2nd}
are employed: (i) a fermionic, and (ii) a qubit representation.
The fermionic Hamiltonian representation of the terms
is denoted 
\begin{equation}
\op{H}_k = h_k \op{F}_k, 
\end{equation}
where
\begin{equation}\label{hermitian_fermionic_ham_terms}
\hat{\mathbf{F}}_k \in
\begin{cases}
a^{\dag}_i a_{i}\\
a^{\dag}_i a_{j} +a^{\dag}_j a_{i} \\
a^{\dag}_p a^{\dag}_q a_{q}a_{p} \\
a^{\dag}_pa^{\dag}_q a_{q}a_{r} + a^{\dag}_r a^{\dag}_q a_{q}a_{p} \\
a^{\dag}_p a^{\dag}_q a_{r}a_{s} + a^{\dag}_s a^{\dag}_r a_{q}a_{p}
\end{cases}.
\end{equation}
It is directly obtained from the second-quantised Hamiltonian
in Eq. \eqref{Ham2nd} \cite{Whitfield2011} by
rearranging the two-body terms, employing the anti-commutation relations
of fermionic operators \cite{HelgakerBible}:
\begin{equation}\label{ferm_grouping_1}
h_{pqrs} \crt{p}\crt{q}\anh{r}\anh{s} + h_{qpsr} \crt{q}\crt{p}\anh{s}\anh{r}=
\left( h_{pqrs}+ h_{qpsr} \right) \crt{p}\crt{q}\anh{r}\anh{s} \ .
\end{equation}
Then, two-body terms involving all-up or all-down spin orbitals,
i.e. $\sigma(p)=\sigma(q)=\sigma(r)=\sigma(s) \in \{ \alpha, \beta \}$,
are sorted,
\begin{equation}\label{ferm_grouping_2}
\begin{aligned}
& \left(h_{pqrs} + h_{qpsr}\right) \crt{p}\crt{q}\anh{r}\anh{s} +
\left( h_{pqsr} + h_{qprs} \right) \crt{p}\crt{q}\anh{s}\anh{r} \\
=& \left( h_{pqrs} + h_{qpsr} - (h_{pqsr} + h_{qprs}) \right) \crt{p}\crt{q}\anh{r}\anh{s} \ .
\end{aligned}
\end{equation}
Finally, non-hermitian terms, such as $h_{ij} \crt{i}\anh{j}$, are joined
with their hermitian conjugate counterpart to obtain the 
$\op{F}_k$ in Eq. \eqref{hermitian_fermionic_ham_terms}.
By contrast,
the qubit Hamiltonian representation $\op{H}_k = h_k \op{P}_k$ is obtained
by applying a fermion-to-qubit mapping to the fermionic operators in Eq. \eqref{Ham2nd}.
Operators $\op{P}_k$ denote Pauli strings, which are Hermitian and unitary operators
corresponding to tensor products of the Pauli (and identity) matrices.

Quantum circuit mappings are known for the unitary operators
$\exp(-i h_k \op{P}_k t)$ required for trotterised simulation with a qubit Hamiltonian \cite{seeley2012bravyi},
as well as the operators $\exp(-i h_k \op{F}_k t)$ required for trotterised
simulation with a fermionic Hamiltonian representation.
The fermionic Hamiltonian exponentials can either be approximated
by applying a fermion-to-qubit mapping to the operators $\op{F}_k$ \cite{Whitfield2011, seeley2012bravyi},
and performing trotterised time evolution of the operator $\exp(-i h_k \op{F}_k t)$.
Alternatively, trotterised simulation with the employed fermionic Hamiltonian
representation could be implemented exactly by utilising fermionic
swap networks \cite{kivlichan2018quantum}.
Arbitrary spin orbitals $p$ and $q$
(for one-body fermionic operators $\op{F}_k$), or $p$, $q$, $r$, $s$
(for two-body fermionic operators $\op{F}_k$)
can be brought to neighbouring qubits by layers of fermionic swap gates.
The time evolution operators $\exp(-i h_k \op{F}_k t)$ can then be exactly
implemented as 2-local, or 4-local qubit unitaries.

We note that more efficient quantum circuit mappings of a fermionic Hamiltonian
representation are achievable by factorisation of
the electron repulsion integral
tensor\cite{MartinezPartitionings, motta2021low, rubin2022compressing}
compared to the fermionic representation we employed in this article.
The focus in this work is, however, on the assessment of different
strategies, which employ orbital transformations in the context
of trotterised Hamiltonian simulation.
The strategies that we investigate should therefore be generally applicable,
independent of the chosen Hamiltonian representation.
As such, the simple fermionic Hamiltonian representation described above
was employed for convenience.

\subsection{Orbital transformations}\label{subsec:orbital_basis_transformations}

The orbital basis, in which the second-quantised Hamiltonian in Eq. \eqref{Ham2nd}
is expressed, can be transformed by unitary transformations $U_R$, which
we refer to as orbital transformations hereafter.
The eigenspectrum of the electronic Hamiltonian is invariant under these
orbital transformations.

Orbital basis transformations can be realised by transformations of
the spatial orbital coefficient matrix $\mathbf{C}$
(see Eq. \eqref{SPATORB} for the definition of the matrix elements
of $\mathbf{C}$) as
\begin{equation}\label{mo_coeff_unitary_transform}
\tilde{\mathbf{C}} = U_{\mathrm{R}}\mathbf{C}.
\end{equation}
Following Ref. \cite{MartinezPartitionings}, arbitrary orbital
transformations can be constructed as products of Givens rotations:

\begin{equation}\label{u_r_via_givens_def}
U_{R} = \prod_{p<q}^{N/2} U_{R, \mathrm{Givens}}(p,q) \ .
\end{equation}
In Eq. \eqref{u_r_via_givens_def}, Givens rotations
$U_{R, \mathrm{Givens}}(p,q)$ are defined as unitary transformations that implement a
pair-wise rotation between the two spatial orbitals at indices $p$ and $q$.
The elements $\tilde{C}_{i, \mu}$ of the transformed coefficient matrix $\tilde{\mathbf{C}}$
resulting from the transformation of the coefficient matrix $\mathbf{C}$
by Eq. \eqref{mo_coeff_unitary_transform} with $U_R$ = $U_{R, \mathrm{Givens}}(p,q)$
are:
\begin{equation}
\begin{aligned}
&\tilde{C}_{p, \mu} = \cos(\theta_{pq}) C_{p, \mu} - \sin(\theta_{pq}) C_{q, \mu} \\
&\tilde{C}_{q, \mu} = \sin(\theta_{pq}) C_{p, \mu} + \cos(\theta_{pq}) C_{q, \mu} \\
&\tilde{C}_{k \neq (p,q) , \mu} = C_{k, \mu} \ , 
&\end{aligned}
\end{equation}
where $\theta_{pq} \in [0, 2\pi)$ denotes the Givens rotation angle.
The transformed coefficient matrix $\tilde{\mathbf{C}}$
defines a set of transformed orthonormal spatial orbitals $\{\tilde{\Phi}_{i}(\mathbf{r})\}_{i=1}^{N/2}$:

\begin{equation}\label{transformed_spatorb}
\tilde{\Phi}_{i} = \left( \sum_{\mu=1}^{N/2} \tilde{C}_{i, \mu} \phi_{\mu}(\mathbf{r}) \right) \ ,
\end{equation}
which differs from the set of spatial orbitals defined in Eq. \eqref{SPATORB}.
Constructing the set of transformed spin orbitals
$\{\tilde{\chi}_{i}(\mathbf{r})\}_{i=1}^{N/2}$ via Eq. \eqref{spinorb_defintion}
leads to transformed one- and two-body integrals $\tilde{h}_{ij}$
and $\tilde{h}_{pqrs}$ from Eqs. \eqref{one_body_integral} and \eqref{two_body_integral}.
The second-quantised Hamiltonian in the transformed basis is then given by

\begin{equation}\label{transformed_ham}
\tilde{\op{H}} =
\sum_{i,j}^N \tilde{h}_{ij}
\crtt{i}\anht{j}
+ \frac{1}{2} \sum_{p,q,r,s}^N \tilde{h}_{pqrs} \crtt{p}\crtt{q}\anht{r}\anht{s} , 
\end{equation}
where $\crtt{i}$ and $\anht{i}$ denote creation and annihilation operators
in the transformed orbital basis, respectively.

Alternatively, the transformation of the second-quantised Hamiltonian can be
realised by transforming the creation and annihilation operators from
the initial basis ($\crt{p}$, $\anh{q}$) to the transformed basis ($\crtt{p}$, $\anht{q}$).
As described in Ref. \cite{HelgakerBible}, a unitary transformation matrix

\begin{equation}\label{U_r_via_ah_exponential}
U_R = \exp(- \mathbf{\kappa}) \ ,
\end{equation}
obtained as the exponential of the matrix representation $\mathbf{\kappa}$
of an anti-hermitian operator $\hat{\kappa} = - \hat{\kappa}^{\dagger}$,
transforms the creation and annihilation operators as

\begin{equation}
\begin{aligned}
&\crtt{p}  = \exp(-\hat{\kappa}) \crt{p} \exp(\hat{\kappa}) \\
&\anht{p}  = \exp(-\hat{\kappa}) \anh{p} \exp(\hat{\kappa}) \ .
\end{aligned}
\end{equation}
To construct the unitary transformation matrix $U_R$ in Eq. \eqref{U_r_via_ah_exponential}
as $U_{R} = U_{R, \mathrm{Givens}}(p,q)$ in the spin orbital basis,
which is needed for a quantum hardware implementation,
we employ the anti-hermitian operator

\begin{equation}\label{givens_rot_kappa_spinorb_basis}
\hat{\kappa} = \hat{\kappa}_{pq} = \theta_{pq} \left( ( \crt{2p-1}\anh{2q-1} - \crt{2q-1}\anh{2p-1} )+
( \crt{2p}\anh{2q} - \crt{2q}\anh{2p} ) \right) \ ,
\end{equation}
where we have mapped spatial orbitals at index $p$
to index $2p-1$ in the spin orbital basis for $\alpha$
spin orbitals and to index $2p$ for $\beta$ spin orbitals.
Inserting Eq. \eqref{givens_rot_kappa_spinorb_basis} into Eq. \eqref{U_r_via_ah_exponential},
the Givens rotation $U_{R, \mathrm{Givens}}(p,q)$ expressed
in the spin orbital basis is then obtained as:
\begin{equation}\label{u_r_givens_spinorb_basis}
U_{R, \mathrm{Givens}}(p,q) = \exp ( -\theta_{pq} ( \crt{2p-1}\anh{2q-1} - \crt{2q-1}\anh{2p-1} )) \cdot
\exp(-\theta_{pq} (\crt{2p}\anh{2q} - \crt{2q}\anh{2p} ) ) \ .
\end{equation}
By expressing the unitary transformation matrix $U_R$ in the spin orbital basis
via Eqs. \eqref{u_r_via_givens_def} and \eqref{u_r_givens_spinorb_basis},
the second-quantised Hamiltonian defined in Eq. \eqref{Ham2nd} can be directly transformed as
\begin{equation}\label{rotation_transform_Ham2nd}
U_{R} \op{H} U_{R}^{\dag}  = \tilde{\op{H}} \ .
\end{equation}

With known quantum circuit decompositions for $U_{R, \mathrm{Givens}}(p,q)$ \cite{arrazola2022universal},
we can hence implement any orbital transformation unitary $U_R$
on a quantum device.

\subsection{Trotter error}\label{subsec:analytical_expressions}

Trotter error describes the error incurred in the approximation of the unitary operator $U=\exp (-i \op{H}t)$
with the trotterised propagator
\begin{equation}\label{H_eff_definition}
U_{\mathrm{Trotter}} = \prod_{k=1}^{\Gamma} \exp(-i\op{H}_k t)  \equiv \exp(-i\op{H}_{\mathrm{eff}} t) \ .
\end{equation}
In Eq. \eqref{H_eff_definition}, $U_{\mathrm{Trotter}}$ is constructed from a
first-order Trotter formula, which we consistently employ hereafter.
Trotterised time evolution effectively corresponds to the exact
Hamiltonian simulation of the effective Hamiltonian $\op{H}_{\mathrm{eff}}=-\ln U_{\mathrm{Trotter}}/(it)$.
Trotter error can be quantified in various ways, (i) as
the difference in the propagator \cite{ChildsTrotter}
\begin{equation}\label{operator_norm_error}
\Delta U (t) = e^{-i\op{H}_{\mathrm{eff}} t} - e^{-i\op{H} t} \ ,
\end{equation}
(ii) as differences in the operator spectrum, such as a
difference in the ground state energies,
\begin{equation}\label{target_eval_error}
\Delta E_0 = E_{0, \mathrm{Trotter}} - E_0,
\end{equation}
(iii) as the difference in the time-evolved Hilbert-space state vectors, measured, for example, as their L$^2$-norm,
\begin{equation}\label{wf_difference}
\lvert \Delta_{\ket{\psi}} \rvert = \lvert e^{-i \op{H}t} \ket{\psi} - 
e^{-i \op{H}_{\mathrm{eff}} t} \ket{\psi}\rvert \ ,
\end{equation}
(iv) as a fidelity measure\cite{FidelityExample}, which is based on the time-evolving overlap
of the true state with the one generated by $\op{H}_{\mathrm{eff}}$,
\begin{equation}\label{fidelity_error}
f(t) = 1 - \braket{\psi(t)}{\psi_\mathrm{Trotter}(t)} = 
1 - \braket{U(t) \psi(0)}{U_{\mathrm{Trotter}}(t) \psi(0)} \ ,
\end{equation}
(v) or to an offset in the autocorrelation function (ACF),
\begin{equation}\label{ACF_offset}
g(t) = | \braket{\psi(0)}{\psi(t)} - \braket{\psi(0)}{\psi(t, \mathrm{Trotter})} | \ .
\end{equation}
We expect the ACF offset $g(t)$ to be strongly correlated with the
Trotter error in the ground state, because the ACF $\braket{\psi(t)}{\psi(0)}$
with $\ket{\psi(0)} = \ket{\psi_0}$ corresponds to a cosine wave that
oscillates with period $2 \pi/E_0$.
We note that upper bounds for the norm of the difference in the propagated wave function,
as defined in Eq. \eqref{wf_difference}, have recently been derived \cite{TrotterStateBounds}.
Since their calculation is computationally demanding, we did not evaluate
them in this work. We refer to Ref. \cite{yi2022spectral}
for a discussion of the relation between fidelity and operator norm errors.
In this work, unless otherwise noted, we 
measure Trotter error by $\Delta E_0$ and refer specifically to the
absolute difference $\lvert \Delta E_0 \rvert$
in the ground state energy of $\op{H}$ and $\op{H}_{\mathrm{eff}}$.

\subsubsection{Analytical error expressions of the propagator and its spectrum}\label{subsubsec:errors_propagator}

While critical for the development of accuracy-guaranteed protocols,
operator spectral norms make the worst-case assumption
$\lVert A \rVert_2 = \sup_{\lvert \vec{v}\rvert = 1} \lvert A \vec{v} \rvert$,
where $A$ denotes an arbitrary bounded operator which acts on elements $\vec{v} \in \mathcal{H}$
of a finite-dimensional Hilbert space $\mathcal{H}$.
For a Trotter step number $s$ in a Hamiltonian simulation with a
guaranteed Trotter error below a targeted accuracy,
an upper bound on $\lvert \Delta E_0 \rvert$ is required.
Lemmata 3 and 4 in the supplementary information of Ref. \cite{Reiher2017}
establish the relation
\begin{equation}\label{relation_operator_norm_dE_Trotter}
\lVert \Delta U(t) \rVert_2 \leq t \gamma(t) \Rightarrow \lVert \op{H}-\op{H}_{\mathrm{eff}} \rVert_2 \leq \gamma(t)
\Rightarrow \lvert \Delta E_0 \rvert \leq \gamma(t)
\end{equation}
between ground-state-energy Trotter error and the 
spectral norm of the propagator difference $\lVert \Delta U(t) \rVert_2$  
under the assumption that $\gamma(t)$ is a non-decreasing
function on the interval $[0, \infty)$.
The currently tightest bounds on $\lVert \Delta U(t) \rVert_2$,
which apply to a general non-sparse Hamiltonian,
were derived by Childs et al. \cite{ChildsTrotter} for
a first-order Trotter formula decomposition,
\begin{align}
&\lVert \Delta U (t) \rVert_2 \leq \frac{t^2}{2} \alpha \label{childs_bound_1st} 
\end{align}
with
\begin{align}
&\alpha = \sum_{a=1}^{\Gamma} {\Big \lVert} \sum_{b>a}^{\Gamma} \left[ \op{H}_b, \op{H}_a \right] {\Big \rVert}_2  \ .\label{childs_bound_alpha_def}
\end{align}
Accuracy-guaranteed protocols are known to overestimate the required Trotter numbers $n$ by multiple
orders of magnitude for electronic Hamiltonians \cite{Reiher2017, BabbushOrbInfluence, mehendale2025estimating, miller2025phase2}.
If the Trotter error is evaluated based on the norm of the propagator
difference, this can, for instance, lead to an incorrect assessment
of low-Trotter-error Hamiltonian partitionings,
as demonstrated in Refs. \cite{MartinezPartitionings, mehendale2025estimating}.
Since the determination of a consistently low Trotter error orbital basis
requires a qualitatively correct numerical
comparison of $\lvert \Delta E_0 \rvert$ across multiple orbital basis,
we require an alternative approximation of the ground-state-energy
Trotter error, rather than upper bounds on Trotter error.

Estimates on ground-state-energy Trotter error from time-independent perturbation theory
have previously been utilised for this purpose \cite{BabbushOrbInfluence}.
Expressions have been derived for first- \cite{MartinezEstimating} and second-order \cite{poulin2015trotter}
Trotter formulae of Hamiltonians in Eq. \eqref{Ham2nd} with an arbitrary
number of terms $\Gamma$.
These expressions derive the leading order
term of the effective Hamiltonian $\op{H}_{\mathrm{eff}}$
by recursive application of the Baker-Campbell-Hausdorff formula.

For the first-order Trotter formula, 
the energy error estimate $\lvert \Delta E_{\mathrm{PT}} \rvert$,
derived in the preprint \cite{MartinezEstimating}, 
reads:
\begin{equation}\label{dE_pt}
\begin{aligned}
\lvert \Delta E_{\mathrm{PT}} \rvert & = | \mel{\psi_0}{\op{H} - \op{H}_{\mathrm{eff}}}{\psi_0} | \\
& = \lvert \mel{\psi_0}{\op{v}_2}{\psi_0} \rvert t^2 +
\sum_{n>0} \frac{| \mel{\psi_n}{ \op{V}_1}{\psi_0}|^2}{E_0 - E_n} t^2 + O(t^3)
\end{aligned}
\end{equation}
with 
\begin{equation}\label{v1_v2_def}
\begin{aligned}
\op{V}_1 & = - \frac{i}{2} \sum_{a = 1}^{\Gamma-1} \sum_{b=a+1}^{\Gamma} [\op{H}_b, \op{H}_a] ,  \\
\op{v}_2 & = - \frac{1}{3} \sum_{a=1}^{\Gamma-1} \sum_{b=a+1}^{\Gamma} \sum_{c=b}^{\Gamma}
\left( 1 - \frac{\delta_{bc}}{2} \right) [\op{H}_c, [\op{H}_b, \op{H}_a]] .
\end{aligned}
\end{equation}
Note that the calculation of $\lvert \Delta E_{\mathrm{PT}} \rvert$ in Eq. \eqref{dE_pt}
is even more cumbersome compared to $\Delta E_0$, since it requires
knowledge of the entire set of eigenvalues $E_n$ and eigenvectors $\ket{\psi_n}$
of the Hamiltonian $\op{H}$.
Ref. \cite{mehendale2025estimating} finds the first and second terms
in Eq. \eqref{dE_pt} to be correlated.
For our analysis of Trotter error for larger molecular systems,
we require a quantity that can be easily calculated and that
correlates with $\lvert \Delta E_0 \rvert$.
Accordingly, we target
\begin{equation}\label{epsilon_2_def}
\epsilon_2 = \lvert \mel{\psi_0}{\op{v}_2}{\psi_0} \rvert
\end{equation}
as a time-independent error estimate from perturbation theory.
Then, $\epsilon_2  t^2$ allows for a comparison with the exact Trotter error
for given times $t$.
In case of poor correlation, an approximation of the second
sum in Eq. \eqref{dE_pt} can always be added as a correction.
However, based on our results, this does not appear to be necessary
for the molecular systems considered in this work.

\subsection{Descriptors correlating with Trotter error}\label{subsec:target_quantities}

Analytical expressions of Trotter error that
are straightforwardly interpretable
in terms of the effect of the orbital basis
are given in Eqs. \eqref{childs_bound_alpha_def} and \eqref{dE_pt}.
But, both equations comprise (nested) commutators of the
Hamiltonian terms $\op{H}_k$ and are therefore not straightforward to evaluate.
However, these equations suggest that the following three descriptors, which depend only on the orbital basis and are easier to evaluate, may be sufficient
to assess Trotter error:
(i) the number of non-negligible coefficient terms $\Gamma$ (because having fewer terms contribute to the (nested) commutator sum should decrease $\alpha$ and $\lVert \op{v}_2\rVert_2$), where we will define a term $\op{H}_k$
as non-negligible, if it features a coefficient weight $\lvert h_k \rvert \leq 10^{-8} E_{\mathrm{h}}$,
(ii) the coefficient weights $\lvert h_k \rvert$,
which are determined for a fermionic Hamiltonian representation
by Eqs. \eqref{hermitian_fermionic_ham_terms}-\eqref{ferm_grouping_2}
and for a qubit Hamiltonian depend on the employed fermion-to-qubit mapping
applied to the operators in Eq. \eqref{Ham2nd},
and (iii) the distribution of the Configuration Interaction (CI) coefficients in the linear combination of
Slater determinants of the ground-state wave function $\ket{\psi_0}$.

Following Eqs. \eqref{mo_coeff_unitary_transform}-\eqref{transformed_ham},
it is apparent that orbital transformations
affect the entire set of one- and two-electron integrals.
As described in section \ref{subsec:ham_rep}, the one- and two-electron
integrals determine the weights $h_k$ of the terms $\op{H}_k$ in the Hamiltonian.
Hence, since the weights $\lvert h_k \rvert$ cannot be minimised independently, one can resort to
minimising their sum $\lambda = \sum_k \lvert h_k \rvert$,
which is usually called the 1-norm for a qubit Hamiltonian representation \cite{OrbMin1norm, qDRIFT, ChristandlRandom2025}.
In general, we will denote
the \enquote{1-norm} as $\lambda_F$ for a fermionic Hamiltonian
and as $\lambda_Q$ for a qubit Hamiltonian.
The relation between upper bounds on Trotter
error in the propagator difference norm and the respective 1-norm 
can be emphasised by deriving an upper bound on
$\alpha$ in Eq. \eqref{childs_bound_alpha_def}.
An upper bound $\alpha' \geq \alpha$ that is directly proportional to $\lambda_F^2$ or $\lambda_Q^2$ is derived in
the Supporting Information (SI).

The influence of the ground state
wave function $\ket{\psi_0}$ on Trotter error
is evident from Eq. \eqref{dE_pt}.
In particular, the multi-reference character
of the ground state wave function might be
of relevance for Trotter error.
This could explain prior findings in Ref. \cite{BabbushOrbInfluence}
that local orbital bases incur large Trotter errors
since $\ket{\psi_0}$ in a local orbital bases can have notable
strong multi-reference character compared to the
canonical basis. Data for our investigated systems which
support this claim can be found in the SI.

\subsection{Construction of randomised bases and series order propagators}\label{subsec:randomized_propagator_deriv}

In this work, we specifically employ propagators,
which change the orbital basis between
Trotter steps.
In trotterised Hamiltonian simulations
with a factorised, fermionic Hamiltonian
representation \cite{MartinezPartitionings},
the orbital basis is transformed within a Trotter step.
For our investigation of the effect of random orbital basis
transformations during trotterised Hamiltonian simulation,
changing the orbital basis between Trotter steps
appears preferable:
it straightforwardly ensures that each term
$\op{H}_k$ is applied in each Trotter step.
Additionally, fewer orbital transformation
unitaries $U_{\mathrm{R}}$ are required, if the basis
is transformed between Trotter steps instead of
(multiple times) inside each Trotter step.
We review the randomised series reordering propagator
introduced by Childs et al.
in Ref. \cite{ChildsRandomization} below.
It matches the one in Ref. \cite{zhang2012randomized}
for a fixed time $t$ per Trotter step.

Starting from the exact propagator $U$ for a given time $t$
defined in Eq. \eqref{TrotterFormula_1}, the time evolution is split into $\eta$ sub-steps,
each evolving the system for equal time $t_n = t/\eta$
\begin{equation}\label{exact_prop_split}
U = \prod_{n=1}^{\eta} \exp(-i\op{H} t_n) \ .
\end{equation}
We now introduce a reordering operator $\mathcal{P}_{\vec{\sigma}}$,
which acts on the Hamiltonian defined in Eq. \eqref{Ham2nd}
by reordering its $\Gamma$ terms according to the elements of the reordering vector $\vec{\sigma}$.
The vectors $\vec{\sigma} \in \mathbb{Z}^{\Gamma}$ are
arrays of shuffled integers from 1 to $\Gamma$, containing each integer
exactly once.
Due to the associativity
of operator addition,
we can therefore insert a different reordering operator
$\mathcal{P}_{\vec{\sigma}_{n}}$
at each time evolution step $n$
into Eq. \eqref{exact_prop_split} to arrive at
the exact expression
\begin{equation}\label{exact_prop_w_reordering}
U = \prod_n \exp(-i \mathcal{P}_{\vec{\sigma}_{n}} \op{H} t_n) \ .
\end{equation}
After Trotterisation of the exponential
operator, Eq. \eqref{trotter_prop_w_reordering} is obtained.
The ordering of the terms in the Hamiltonian now affects the resulting propagator \cite{TrotterOrderingImpact}
and hence the spectrum of the effective Hamiltonian
\begin{equation}\label{trotter_prop_w_reordering}
U_{\mathrm{reo}} = \prod_n \prod_{k}^{\Gamma} \exp(-i \op{H}_{k(\vec{\sigma}_{n})} t_n) \ .
\end{equation}
To emphasise the dependence of the term index $k$ on the explicit reordering in each step $n$, we mark the index
$k$ as being functionally dependent on the reordering vector $\vec{\sigma}_{n}$.

Analogously, the derivation of the trotterised orbital basis rotation propagator
$U_{\mathrm{rot}}$ also starts from Eq. \eqref{exact_prop_split}.
We begin by inserting the identity operator
$\mathcal{I} = U_{R,n}^{\dag} U_{R,n}$ to obtain

\begin{equation}\label{exact_prop_w_ident_insert}
U = \prod_n U_{R,n}^{\dag} U_{R,n} \exp(-i\op{H} t_n) U_{R,n}^{\dag} U_{R,n} \ .
\end{equation}
The orbital transformations $U_{R,n}$ expressed in
the spin orbital basis are defined here following Eqs. \eqref{u_r_via_givens_def}
and \eqref{u_r_givens_spinorb_basis}.
The subscript $n$ again denotes that the orbital
transformations may vary from time evolution step to time evolution step.
Inserting the identity

\begin{equation}\label{U_expA_Udag_identity}
\exp(U_R \op{H} U_R^{\dag}) = U_R \exp(\op{H}) U_R^{\dag}
\end{equation}
into Eq. \eqref{exact_prop_w_ident_insert}, we arrive at the expression

\begin{equation}\label{exact_prop_U_into_exp}
U = \prod_n U_{R,n}^{\dag} \exp(-i U_{R,n} \op{H} U_{R,n}^{\dag} t_n) U_{R,n} \ .
\end{equation}
With Eq. \eqref{rotation_transform_Ham2nd}, we then obtain
the exact propagator

\begin{equation}\label{exact_prop_w_rot}
U = \prod_n U_{R,n}^{\dag} \exp(-i \tilde{\op{H}}^{(n)}  t_n) U_{R,n} \ ,
\end{equation}
where $\tilde{\op{H}}^{(n)}$ denotes the second-quantised Hamiltonian
in the respective orbital basis of the $n$'th time evolution step.
The trotterised propagator, which may become subject to benefit from error cancellation
effects induced by a transformation of the orbital basis, is finally obtained as

\begin{equation}\label{trotter_prop_w_rot}
U_{\mathrm{rot}} = \prod_n U_{R,n}^{\dag} \left( \prod_{k}^{\Gamma} \exp(-i \tilde{\op{H}}^{(n)}_{k}  t_n) \right) U_{R,n} \ .
\end{equation}
We employ the following notation to differentiate between the \enquote{$\eta$-basis-propagators} $U_{\mathrm{rot}}$ and the \enquote{$\eta$-ordering-propagators}
$U_{\mathrm{reo}}$, and their \enquote{constituent} propagators,
$U_{\mathrm{rot},n}$ and $U_{\mathrm{reo},n}$:

\begin{equation}\label{U_rot_def}
U_{\mathrm{rot}} = \prod_n U_{R,n}^{\dag} U_{\mathrm{rot}, n} U_{R,n} \
\end{equation}
with 
\begin{equation}\label{U_rot_n_def}
U_{\mathrm{rot}, n} =  \prod_{k}^{\Gamma} \exp(-i \tilde{\op{H}}^{(n)}_k  t_n) \ ,
\end{equation}
and
\begin{equation}\label{U_reo_def}
U_{\mathrm{reo}} = \prod_n U_{\mathrm{reo}, n} \
\end{equation}
with
\begin{equation}\label{U_reo_n_def}
U_{\mathrm{reo}, n} = \prod_{k}^{\Gamma} \exp(-i \op{H}_{k(\vec{\sigma}_{n})} t_n) .
\end{equation}
The propagators $U_{\mathrm{rot}}$ and $U_{\mathrm{reo}}$
evolve the wave function for a total time $t$, whereas
the constituent propagators $U_{\mathrm{rot},n}$ and $U_{\mathrm{reo},n}$ evolve the wave function for a time step $t_n$.

Additionally, $\vec{\theta}$ will be used as a
short-hand notation for the Givens rotation representation of the orbital-basis unitary transformations $U_{R}$
as defined in Eq. \eqref{u_r_via_givens_def}.
The elements of the vector $\vec{\theta} \in \mathbb{R}^{\left(\frac{N}{2} \left( \frac{N}{2}-1) \right) /2 \right)}$
list the coefficients $\theta_{pq}$ of the individual Givens rotations
$U_{R,\mathrm{Givens}}(p,q)$ (see Eq. \eqref{u_r_givens_spinorb_basis})
from which $U_R$ is constructed. 
We employ the canonical orbital basis as the reference basis
which defines the zero-vector $\vec{\theta} = \vec{0}$.

We finally note that in the context of Hamiltonian simulation
for a time $t_{\mathrm{total}}$, the $\eta$-basis-propagators
and $\eta$-ordering-propagators introduced in this section
are akin to periodically driven quantum systems \cite{trivedi2025noise}.
This is because in Eq. \eqref{U_rot_def} and \eqref{U_reo_def}
we defined that $U_{\mathrm{rot}}$ and $U_{\mathrm{reo}}$
evolve the wave function for a total time $t$, which is the time
evolution associated with a single Trotter step
and not the total time evolution $t_{\mathrm{total}}$
of Hamiltonian simulation (see Eq. \eqref{TrotterFormula_1}).
A Hamiltonian simulation of time $t_{\mathrm{total}}$
with a Trotter step number $s$ as defined by Eq. \eqref{TrotterFormula_1}
corresponds to a trotterised simulation with a Trotter
step number $s\cdot \eta$, where the orbital basis is
changed every Trotter step if an $\eta$-basis-propagator
is employed.
If the same $\eta$-basis propagator is continuously
applied, the orbital basis changes periodically
(with a period of $\eta$ Trotter steps)
throughout the Hamiltonian simulation.
To obtain randomised propagators which randomly vary
the orbital basis each Trotter step hence requires
the choice of a Trotter step number $s=1$,
based on the definitions employed throughout this work.
This brief excursion to periodically driven quantum systems
hence highlights that the idea to utilise \enquote{modulation schemes}
such that the simulated effective Hamiltonians reproduce
exact Hamiltonian simulation more accurately \cite{goldman2014periodically}
is not unique to randomised propagator constructions.

\section{Computational methodology}\label{sec:numerical_details}

All Cartesian coordinates of the molecular structures were taken from
PubChem \cite{PubChem2019}, with the exception the cyclobutadiene (CBD) molecule,
whose structure we optimised within a Kohn-Sham DFT calculation with the
geomeTRIC library \cite{geomeTRIC}
using the B3LYP exchange-correlation energy functional \cite{b3lyp}
and a cc-pVTZ basis set \cite{cc_basis_sets}.

For the few-atomic systems investigated in
section \ref{subsec:results_atoms_dimers}, we chose the minimal sto-3g basis set \cite{hehre1969self}
to obtain results, which can be
compared to those in Ref. \cite{BabbushOrbInfluence}.
Active spaces are chosen around the Fermi level when employing a
canonical orbital basis.
For restricted MP2 (RMP2) natural orbitals,
the equivalent to the Fermi level is placed at the threshold, where
the spatial orbital occupancy drops sharply from close to two (occupied orbitals)
to close to zero (virtual orbitals).
These active spaces contain all orbitals except the frozen core orbitals.
The number of electrons in the active space relates to the number of occupied Hartree-Fock or RMP2 orbitals, assigning all other orbitals in the active space to the virtual orbital space; the resulting active spaces 
are listed in Table \ref{tab:system_oview}.

For small molecular $\pi$-systems, we employed the cc-pVTZ basis set \cite{cc_basis_sets},
for the $\pi$-system of octatetraene and larger ones
the cc-pVDZ basis set \cite{cc_basis_sets} was employed.
The active space spatial orbitals corresponding to
the $\pi$-orbitals were selected manually
as follows:
For $\pi$-systems with $N/4$ conjugated $\pi$ bonds,
the selected active spaces of $N/2$ spatial orbitals
and $M={N}/{2}$ electrons include $N/4$ spatial orbitals that are occupied
in the HF (or RMP2 natural orbital) reference wave function
and $N/4$ spatial orbitals from the virtual space. The selected occupied orbitals were visually identified
as $\pi$-bonding orbitals and the virtual orbitals as their
anti-bonding counterparts.
For the pyrrole molecule with $N/4=2$ conjugated $\pi$-bonds,
an additional spatial orbital from the occupied space that corresponded to a $p_z$-type orbital was added to the active space.
Exact details on the employed active spaces are provided in the
Zenodo repository that accompanies this work.

Orbital localisations were carried out with the Foster-Boys scheme \cite{FosterBoys} as
implemented in PySCF \cite{sun2018pyscf}.
Restricted MP2 (RMP2) natural orbitals were obtained with PySCF.
As an example, $\pi$-orbitals in different bases are shown for butadiene
and naphthalene in the SI.

The second-quantised electronic Hamiltonian was obtained
in a spin orbital basis with the function \enquote{qchem.fermionic\_observable}
implemented in the PennyLane python library \cite{Pennylane2018}.
The one- and two-electron integrals
were calculated with PySCF.
As previously described in section \ref{subsec:target_quantities},
we neglected all terms $\op{H}_k$ with coefficient weights $\lvert h_k \rvert$ below $10^{-8}$ $E_{\mathrm{h}}$.
The qubit Hamiltonian was obtained by
Jordan-Wigner mapping \cite{JWOriginal} of the electronic Hamiltonian, as implemented in PennyLane.

Two orderings of the Trotter series were employed in this
work: (i) a magnitude ordering \cite{TrotterOrderingImpact},
where the terms $\op{H}_k$ were ordered by decreasing $\lvert h_k \rvert$ to
incur consistently low Trotter error independent of the orbital basis,
and (ii) index-dependent orderings for fermionic and qubit
Hamiltonian representations, which
facilitate the evaluation of the effect of orbital
transformations on Trotter error with fixed Trotter series-orderings.
Details on utilised index-dependent orderings are provided in the SI.

The exact Trotter energy error was calculated in two ways:
up to (but excluding) section \ref{subsec:error_cancellation_numerical_display}
$\Delta E_0$ was calculated by inverse Fourier transform (IFT) of
the ACF as
\begin{equation}\label{dE_to_precision_epsilon}
\Delta E_0 = E_0(\epsilon) - E_{0, \mathrm{Trotter}}(\epsilon) \ ,
\end{equation}
where
\begin{equation}\label{E_0_via_IFT}
E_{0}(\epsilon) = \min \left( \mathrm{IFT \left( \braket{\psi_0}{\psi(t)} \right)  }  \right)
\end{equation}
and
\begin{equation}\label{E_0_Trotter_via_IFT}
E_{0, \mathrm{Trotter}}(\epsilon) = \min \left( \mathrm{IFT \left( \braket{\psi_0}{\psi_{\mathrm{Trotter}}(t)} \right)  }  \right)
\end{equation}
with
$\ket{\psi(t)}$ and $\ket{\psi_{\mathrm{Trotter}}(t)}$ defined in Eq. \eqref{fidelity_error}.
Eigenenergies of the Hamiltonian by IFT of
the ACF can only be obtained up to precision $\epsilon$,
depending on the total time evolution length $t_{\mathrm{total}}$.

For a comparison of Trotter errors of $\eta$-basis-propagators and
$\eta$-ordering-propagators (Eqs. \eqref{U_rot_def} and \eqref{U_reo_def})
and their individual constituents (Eqs. \eqref{U_rot_n_def} and \eqref{U_reo_n_def}),
we employed a second, more convenient approach, where
we calculated Trotter error $\Delta E_0$ by diagonalisation of
the spin-orbital matrix representation of
the effective Hamiltonian $\op{H}_{\mathrm{eff}}$
directly, as defined in Eq. \eqref{H_eff_definition}.
This is beneficial as
we were not restricted with respect to the precision $\epsilon$
with which we would obtain $E_0$ and $E_{0, \mathrm{Trotter}}$
due to a finite-length ACF.

The ground state wave function $\ket{\psi_0}$ was obtained
from Density Matrix Renormalization Group (DMRG)-CI
calculations with the QCMaquis computer program \cite{szenes2025qcmaquis} for
a bond dimension of 1024.
To map the ground-state matrix-product-state wave function to a sum of occupation
number vectors in the determinantal basis,
we exploited the SR-CAS algorithm \cite{boguslawski2011construction}
as implemented in QCMaquis \cite{szenes2025qcmaquis}.

Our construction of the trotterised and exact propagators
$U_{\mathrm{Trotter}}(t)$ and $U(t)$, respectively, utilised
the PennyLane \cite{Pennylane2018}, OpenFermion \cite{openfermion2020},
and scipy \cite{scipy2020} libraries.
Parts of the trotterised propagator construction utilise
the code repository \cite{githubMartinez} of Ref. \cite{mehendale2025estimating}.
Unless otherwise noted, Trotter step numbers were chosen
such that $t = 0.95 {\pi}/{\lvert E_0 \rvert}$.
This choice of step size assured that we
were using the maximum possible step size while adhering
to the Nyquist-Shannon sampling theorem \cite{ShannonSampling}
for the Fourier transformation.
The factor of $0.95$ guarantees
that the signal $E_{0, \mathrm{Trotter}}$
will not be cut off in the spectrum if $E_{0, \mathrm{Trotter}}$ is smaller than $E_0$.
We refer to Table \ref{tab:system_oview} for an overview of the resulting times $t$ for
the systems considered in this work.

A python program was written for the calculation of
the upper bound of the Trotter error $\alpha$
and for the perturbative error estimate $\epsilon_2$
(according to Eq. \eqref{childs_bound_alpha_def} and Eq. \eqref{epsilon_2_def}, respectively). 

For commutator elements of
the fermionic Hamiltonian representation,
we calculated
$[\op{F}_b, \op{F}_a]$ with an in-house program by implementing the analytical
results of the commutators between normal-ordered one- to
three-electron terms (see the SI for explicit expressions).
For commutator elements of the qubit
Hamiltonian representation,
the PennyLane library was exploited to calculate
commutators of Pauli strings.

Moreover, the calculation of $\epsilon_2$
requires the evaluation of up to four-electron operator
matrix elements.
We used QCMaquis \cite{szenes2025qcmaquis} to calculate the unique elements
of all non-zero one-body to four-body reduced density matrices (one- to 4-RDMs)
in the spin orbital
basis. Obviously, the calculation of 4-RDMs is expensive
due to the number of elements, scaling with the number of active space
orbitals as $N^8$.
This steep scaling creates a natural restriction to
up to 10 spatial orbitals (20 spin orbitals, and hence 20 qubits) to calculate
estimates of the Trotter energy error $\epsilon_2$.
Even for the 10 spatial orbital case,
owing to the sheer number of terms in the Hamiltonian
$\op{H}_k$, calculating the double commutator sum $\op{v}_2$
is very expensive despite the calculation of
$\op{v}_2$ in Eq. \eqref{v1_v2_def} being embarrassingly parallel.

Calculating $\epsilon_2$
for a qubit Hamiltonian representation required
the double commutator sum of Pauli strings to be evaluated.
Afterwards, we employed the reverse Jordan-Wigner mapping
implemented in OpenFermion \cite{openfermion2020}
to map the resulting sum of Pauli strings to a sum
of normal-ordered fermionic operators.
The previously calculated one- to four-body RDM elements were
then exploited to calculate $\epsilon_2$.

We calculated the fidelity error $f(t)$ and the autocorrelation
function offset $g(t)$ after a single time step
with $t=0.95 {\pi}{/\lvert E_0 \rvert}$.
We observed the best correlations with exact-error
calculations after a single step, when compared to calculations after multiple time steps.
The calculation of $f(t)$ and $g(t)$ was performed by the tangent-space time-dependent DMRG algorithm \cite{baiardi2019large}
as implemented in QCMaquis \cite{szenes2025qcmaquis}.
In a local QCMaquis version, the construction of the
trotterised time evolution matrix product operators $e^{-i h_k \op{F}_kt}$ was implemented.

Orbital transformations were obtained as products of Givens rotations
following Eqs. \eqref{u_r_via_givens_def} and \eqref{u_r_givens_spinorb_basis} by direct transformation
of creation and annihilation operators:
\begin{equation}\label{crt_transform_giv_rot}
\begin{pmatrix}
\crtt{p} \\ \crtt{q}
\end{pmatrix}
 = 
\begin{pmatrix}
\cos \theta_{pq} & -\sin \theta_{pq} \\
\sin \theta_{pq} &  \cos \theta_{pq}
\end{pmatrix}
\begin{pmatrix}
\crt{p} \\ \crt{q}
\end{pmatrix}
\end{equation}
and
\begin{equation}\label{anh_transform_giv_rot}
\begin{pmatrix}
\anht{p} \\ \anht{q}
\end{pmatrix}
 = 
\begin{pmatrix}
\cos \theta_{pq} & -\sin \theta_{pq} \\
\sin \theta_{pq} &  \cos \theta_{pq}
\end{pmatrix}
\begin{pmatrix}
\anh{p} \\ \anh{q}
\end{pmatrix} \ .
\end{equation}
Consequently, the creation and annihilation operators
can be directly transformed to obtain $\tilde{\op{H}}^{(n)}$
from $U_{R,n}$ and $\op{H}$ according to Eq. \eqref{rotation_transform_Ham2nd}
without the explicit construction of the matrix $U_{R,n}$
in the spin orbital basis.

To allow for a straightforward comparison of Trotter error $\Delta E_0$
of propagators $U_{\mathrm{reo}}$ and $U_{\mathrm{reo}, n}$
(as well as $U_{\mathrm{rot}}$ and $U_{\mathrm{rot}, n}$)
obtained for different molecules, we normalised the spectra
of their Hamiltonians and shifted them to the interval $[-1, 0]$.

All plots in this work were created with the matplotlib python package \cite{hunter2007matplotlib}.

\section{Results}\label{sec:results}

In this section, we present
results for the fermionic Hamiltonian representation,
since it enabled the analysis of the additional estimates of Trotter
error $f(t)$ and $g(t)$ (see Eqs. \eqref{fidelity_error} and \eqref{ACF_offset})
as described in section \ref{sec:numerical_details}.
The results for the qubit Hamiltonian representation are provided in the SI.
However, a discussion of differences and similarities in the results obtained for
the two different Hamiltonian representations
has been included in this section.
Section \ref{subsec:results_atoms_dimers} discusses
Trotter errors for atoms in various orbital bases similar to those in Ref. \cite{BabbushOrbInfluence}.
In section \ref{subsec:results_pi_sys}, we compare Trotter error
for $\pi$-systems with up to six spatial orbitals
in different orbital basis sets.
We evaluate the correlation of Trotter error estimates
$\alpha t^2$, $\epsilon_2 t^2$, $g(t)$ and $f(t)$
with exact ground state energy Trotter error $\lvert \Delta E_0 \rvert$.
For estimates which show notable correlation with Trotter error,
we perform calculations on larger $\pi$-systems
with active spaces of up to ten spatial orbitals.
We report on the correlation of descriptors with Trotter
error in \ref{subsubsec:results_med_pi_target_quant_correlation}.
Randomised $\eta$-basis- and $\eta$-ordering-propagators (see Eqs. \eqref{U_rot_def} and $\eqref{U_reo_def}$)
are discussed in section \ref{subsec:error_cancellation_numerical_display}.
Finally, we investigated whether the observed continuity of Trotter error
as a function of the transformed orbital basis $\vec{\theta}$
allows for the determination of an orbital basis without a
Trotter error by a bisection method in the SI.

\subsection{Atomic, diatomic, and triatomic systems}\label{subsec:results_atoms_dimers}

\begin{figure}[H]
    \centering
    \includegraphics[width=0.85\linewidth]{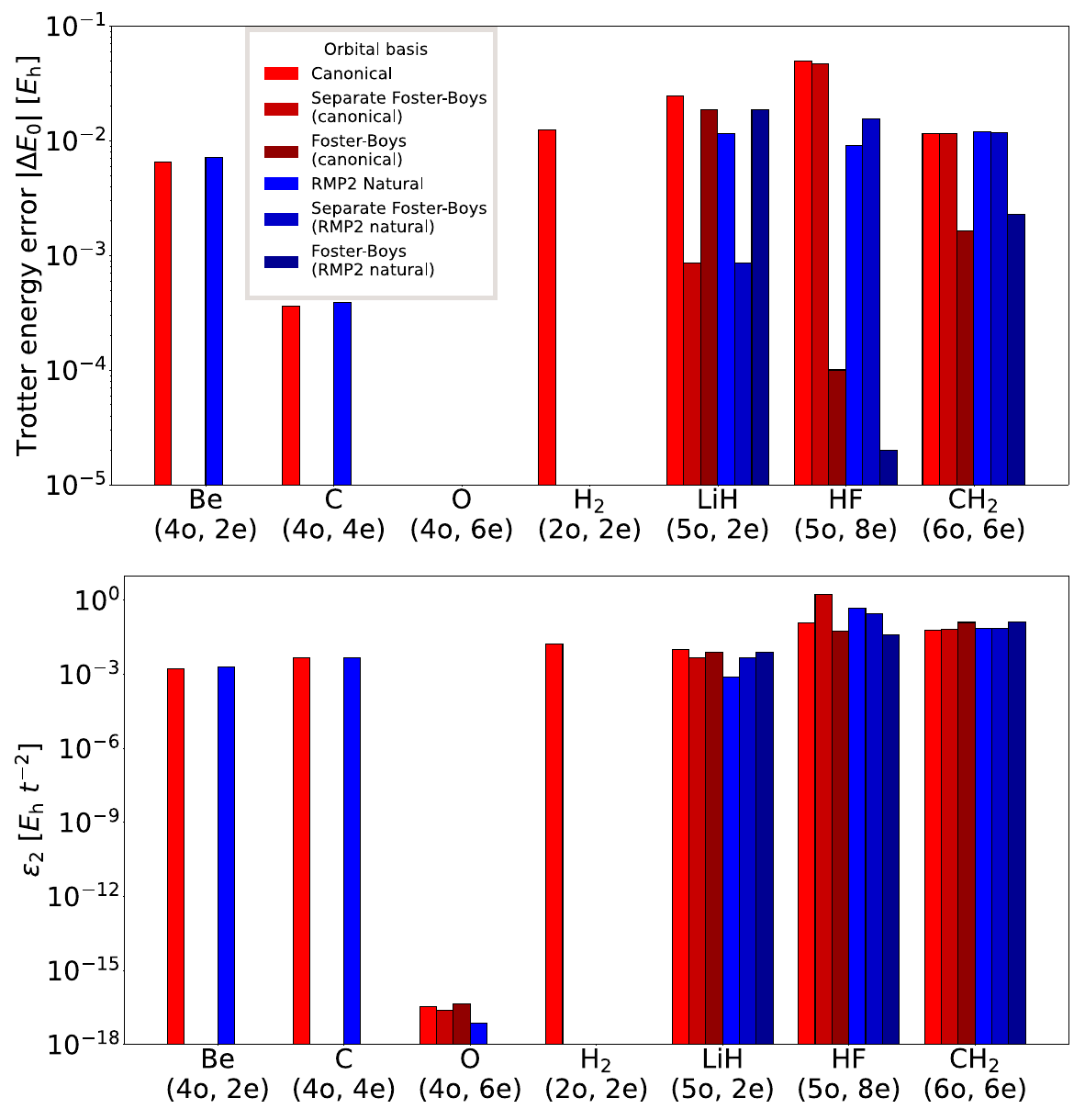}
    \caption{Top: Trotter energy errors $\lvert \Delta E_0 \rvert$
    obtained to precision $\epsilon=10^{-5}$ $E_{\mathrm{h}}$ (see Eqs. \eqref{dE_to_precision_epsilon}-\eqref{E_0_Trotter_via_IFT}) in various common orbital bases.
    Bottom: perturbation theory energy error estimate
    $\epsilon_2$ defined in Eq. \eqref{epsilon_2_def}.
    Results were obtained for small atomic systems
    in a sto-3g basis.
    For evolution times $t$ per Trotter step see
    Table \ref{tab:system_oview}.
    The fermionic Hamiltonian representation
    and a magnitude ordering of the Trotter series were employed.}
    \label{fig:small_orb_trotter_errors}
\end{figure}

Fig. \ref{fig:small_orb_trotter_errors} shows
how the Trotter energy error depends on the choice of
orbital basis for small systems of denoted active space
sizes.
The following six orbital bases were considered:
the canonical orbital basis, a separately localised
orbital basis starting from canonical orbitals,
and localised orbitals starting from canonical orbitals.
The RMP2 natural orbital basis, a separately localised
orbital basis starting from RMP2 natural orbitals,
and localised orbitals starting from RMP2 natural orbitals.
A separately localised canonical (or RMP2 natural) orbital basis refers to
the orbital basis obtained from separate orbital localisation of the occupied
and virtual orbitals in the canonical (or RMP2 natural) orbital basis active space.
A localised canonical (or RMP2 natural) orbital basis refers to
the orbital basis obtained from a single orbital localisation, which includes,
and, hence, \enquote{mixes} the occupied and virtual orbitals in the
canonical (or RMP2 natural) orbital basis active space.

We have omitted (separately) localised orbital bases data from
the plot for the few-atomic systems (Be, C, O, H$_2$), where
the Foster-Boys orbital localisation was inapplicable
(see the missing bars for the Be and C atom systems
in Fig. \ref{fig:small_orb_trotter_errors} for example).
For molecular hydrogen ($\mathrm{H_2}$), the canonical and RMP2 natural
orbitals are identical, which is why the RMP2 natural orbital basis
result is omitted in Fig. \ref{fig:small_orb_trotter_errors}.
Active spaces of ${N}/{2}$ spatial orbitals and $M$ electrons
are denoted as $({N}/{2}\mathrm{o}, \ M\mathrm{e})$.
The evolution times $t$ employed in the construction of $U_{\mathrm{Trotter}}(t)$
are provided in Table \ref{tab:system_oview}.

\begin{table}[h]
    \centering
    \begin{tabular}{c|c|c|c|c|c}\hline\hline
         Molecule & basis set & CAS & CAS &
         CASCI  & Time per Trotter \\
         &  & orbitals &  electrons & energy [$E_{\mathrm{h}}$] & step $t$ $[E_{\mathrm{h}}^{-1}]$\\
         \hline
         $\mathrm{Be}$ atom& sto-3g & 4 & 2 & -0.978 & 3.051 \\
         $\mathrm{C}$ atom& sto-3g & 4 & 4 & -5.247 & 0.569 \\
         $\mathrm{O}$ atom& sto-3g & 4 & 6 & -15.367 & 0.194 \\
         $\mathrm{H_2}$ & sto-3g & 2 & 2 & -1.851 & 1.612 \\
         $\mathrm{LiH}$ & sto-3g & 5 & 2 & -1.080 & 2.762 \\
         $\mathrm{HF}$ & sto-3g & 5 & 8  & -28.020 & 0.107 \\
         $\mathrm{CH_2}$ & sto-3g & 6 & 6 & -10.659 & 0.28 \\
         \hline
         $\mathrm{butadiene}$ (BD)& cc-pVTZ & 4 & 4 & -3.308 & 0.902 \\
         $\mathrm{cyclobutadiene}$ (CBD)& cc-pVTZ & 4 & 4 & -3.519 & 0.848 \\
         $\mathrm{pyrrole}$ (Pyrr)& cc-pVTZ & 5 & 6 & -7.044 & 0.424 \\
         $\mathrm{hexatriene}$ (Hexa)& cc-pVTZ & 6 & 6 & -5.834 & 0.512 \\
         $\mathrm{benzene}$ (Benz)& cc-pVTZ & 6 & 6 & -6.442 & 0.463 \\
         $\mathrm{pyridine}$ (Pyri)& cc-pVTZ & 6 & 6 & -6.756 & 0.442 \\
         $\mathrm{octatetraene}$ (Octa)& cc-pVDZ & 8 & 8 & -11.207 & 0.266 \\
         $\mathrm{decapentaene}$ (Deca)& cc-pVDZ & 10 & 10 & -11.671 & 0.256 \\
         $\mathrm{naphthalene}$ (Naph)& cc-pVDZ & 10 & 10 & -13.795 & 0.216 \\
         \hline\hline
    \end{tabular}
    \caption{Systems and their abbreviations in parentheses,
    their associated active space sizes (number of spatial orbitals $N/2$ and number of electrons $M$),
    and ground state energy $E_0$ of the Complete Active Space (CAS)-CI
    Hamiltonian for the RMP2 natural orbitals active space.
    The time evolution per Trotter step was determined as $t = 0.95 {\pi}/{\lvert E_{0} \rvert}$.
    In case of ambiguity, the names for linear conjugated $\pi$-systems refer to the \enquote{all} (\textit{E}) isomer,
    that is, the IUPAC name for the selected octatetraene isomer would be (3\textit{E},5\textit{E})-octa-1,3,5,7-tetraene.}
    \label{tab:system_oview}
\end{table}

The Trotter error $\lvert \Delta E_0 \rvert$ for the oxygen atom is below
the energy-resolution threshold of $\epsilon=10^{-5} \ E_{\mathrm{h}}$
for these calculations, and can hence not be shown in Fig. \ref{fig:small_orb_trotter_errors}.
The differing values of $\epsilon_2$ for the oxygen atom
are a result of numerical error, due to the very small values for
$\lvert \epsilon_2 \rvert$ which
are below $10^{-16} \ E_{\mathrm{h}}\  t^{-2}$.

It is obvious from Fig. \ref{fig:small_orb_trotter_errors}
that the choice of orbital basis can significantly affect
Trotter error.
The most notable example is observed for the
$\mathrm{HF}$ molecule, where the energy error $\lvert \Delta E_0 \rvert$
incurred due to Trotterisation can vary by up to 3 orders of
magnitude between different orbital bases.
Similar results are observed for a qubit Hamiltonian
representation which we provide in the SI.
We note that when employing the qubit Hamiltonian representation,
the Trotter error of the oxygen atom deviates
notably less compared to the other systems.

Similar to Ref. \cite{BabbushOrbInfluence}, we find
for a perturbative error estimate of Trotter error
that the oxygen atom incurs a Trotter error, which is
many orders of magnitude smaller compared to other
investigated systems.
We observe comparable results regarding the variation of
a Trotter error estimate with the orbital basis choice to
Ref. \cite{BabbushOrbInfluence}.
Namely, the difference in Trotter errors between different systems
is of comparable magnitude, and the variance in error with different
orbital choices are generally in agreement.
Differences in our results are to
be expected, given the use of a different order Trotter
formula (first versus second order),
different Trotter series-ordering
(magnitude \cite{TrotterOrderingImpact} 
versus lexicographic
ordering \cite{TrotterLexicographicOrder, BabbushOrbInfluence})
and slightly different active spaces (the core orbitals are frozen in our case,
whereas they are included in the active space in Ref. \cite{BabbushOrbInfluence}).

Ref. \cite{BabbushOrbInfluence} reported that the CASCI natural orbital
basis or canonical basis lead to reduced Trotter error compared to a local basis
in the majority of cases.
We did not observe a clear trend
which choice of orbital basis generally leads to the lowest Trotter errors, because of
(i) the increased number of investigated orbital bases
compared to Ref. \cite{BabbushOrbInfluence} per molecule, and
(ii) the limited number of available data points for localised bases
of the atomic systems for which (Foster-Boys) orbital
localisation is inapplicable.

\subsection{\texorpdfstring{$\pi$}{π}-conjugated molecules}\label{subsec:results_pi_sys}

\begin{figure}[H]
    \centering
    \includegraphics[width=0.85\linewidth]{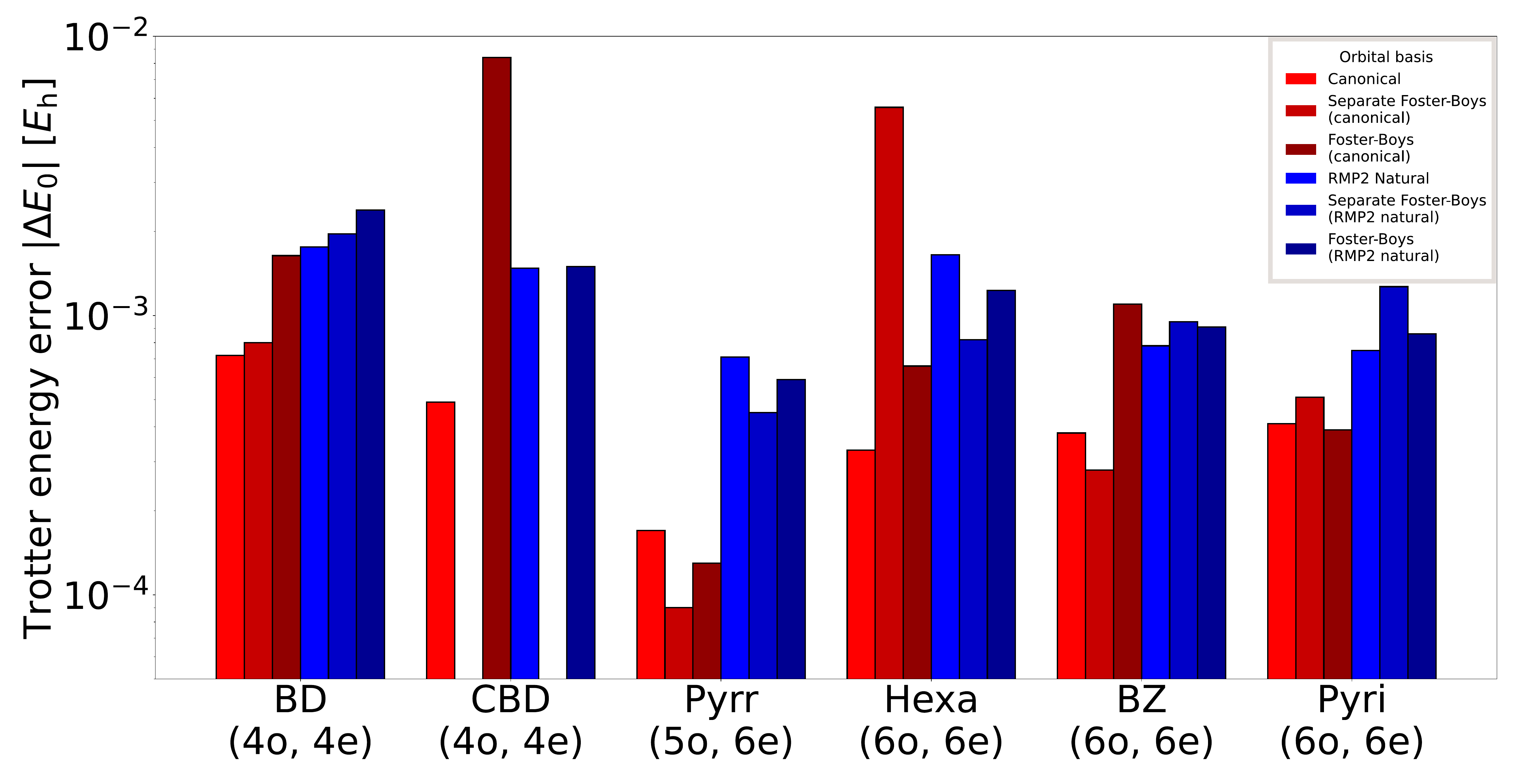}
    \caption{Trotter energy error $\lvert \Delta E_0 \rvert$ obtained 
    to precision $\epsilon=10^{-5}$ $E_{\mathrm{h}}$ (see Eqs. \eqref{dE_to_precision_epsilon}-\eqref{E_0_Trotter_via_IFT})
    for $\pi$-systems with active spaces of up to 6 spatial orbitals in various common orbital bases.
    The fermionic Hamiltonian representation
    and a magnitude ordering of the Trotter series were employed.
    For evolution times $t$ per Trotter step
    and for the abbreviations of the different molecules,
    we refer to Table \ref{tab:system_oview}.}
    \label{fig:med_orb_vs_dE}
\end{figure}

We now investigate Trotter errors for a set of homologous
molecules, where the effect of orbital localisation
is more pronounced:
orbital active spaces comprising the $\pi$-electrons
in the $\pi$-systems.
Based on the data tabulated in the SI, we found:
(i) The difference between canonical and local basis
(measured by differences in the 1-norm)
for $\pi$-systems versus atomic systems is increased, 
(ii) the orbitals in $\pi$-system active spaces retain
a consistent character with an increasing number of ethylene
units, and (iii) the multi-reference character of $\ket{\psi_0}$ increases with
the number of conjugated $\pi$ bonds in these systems \cite{kurashige2014theoretical, bettinger2016electronically}, both for
canonical and localised bases.
Trotter energy errors $\lvert \Delta E_0 \rvert$ for the $\pi$-systems
with up to six spatial orbitals are shown in Fig. \ref{fig:med_orb_vs_dE} (abbreviations of the names of the $\pi$-systems
are defined in Table \ref{tab:system_oview}).
For the CBD system, the separate Foster-Boys localisation, starting both from canonical and
natural orbitals, did not converge. Hence, these data points were omitted in
Fig. \ref{fig:med_orb_vs_dE}.

At most, up to an order of magnitude variance in Trotter error is observed
for each system in Fig. \ref{fig:med_orb_vs_dE}.
In most cases, Trotter errors for (localised) canonical orbitals
are decreased compared to (localised) RMP2 natural orbitals.
The decrease is minor and might simply be a
consequence of the slightly smaller CASCI energies associated with
the RMP2 orbital active spaces compared to the canonical orbital
active spaces (see the SI for details).
Results for the qubit Hamiltonian representation
are shown in the SI.
A similar variance of up to an order of magnitude
in Trotter error is observed among the investigated orbital bases.
The data does not suggest a consistent effect of
orbital localisation on Trotter error.

\subsubsection{Comparison of exact Trotter error and error estimates}\label{subsubsec:results_med_pi_estimate_evaluation}

\begin{figure}[H]
    \centering
    \includegraphics[width=0.85\linewidth]{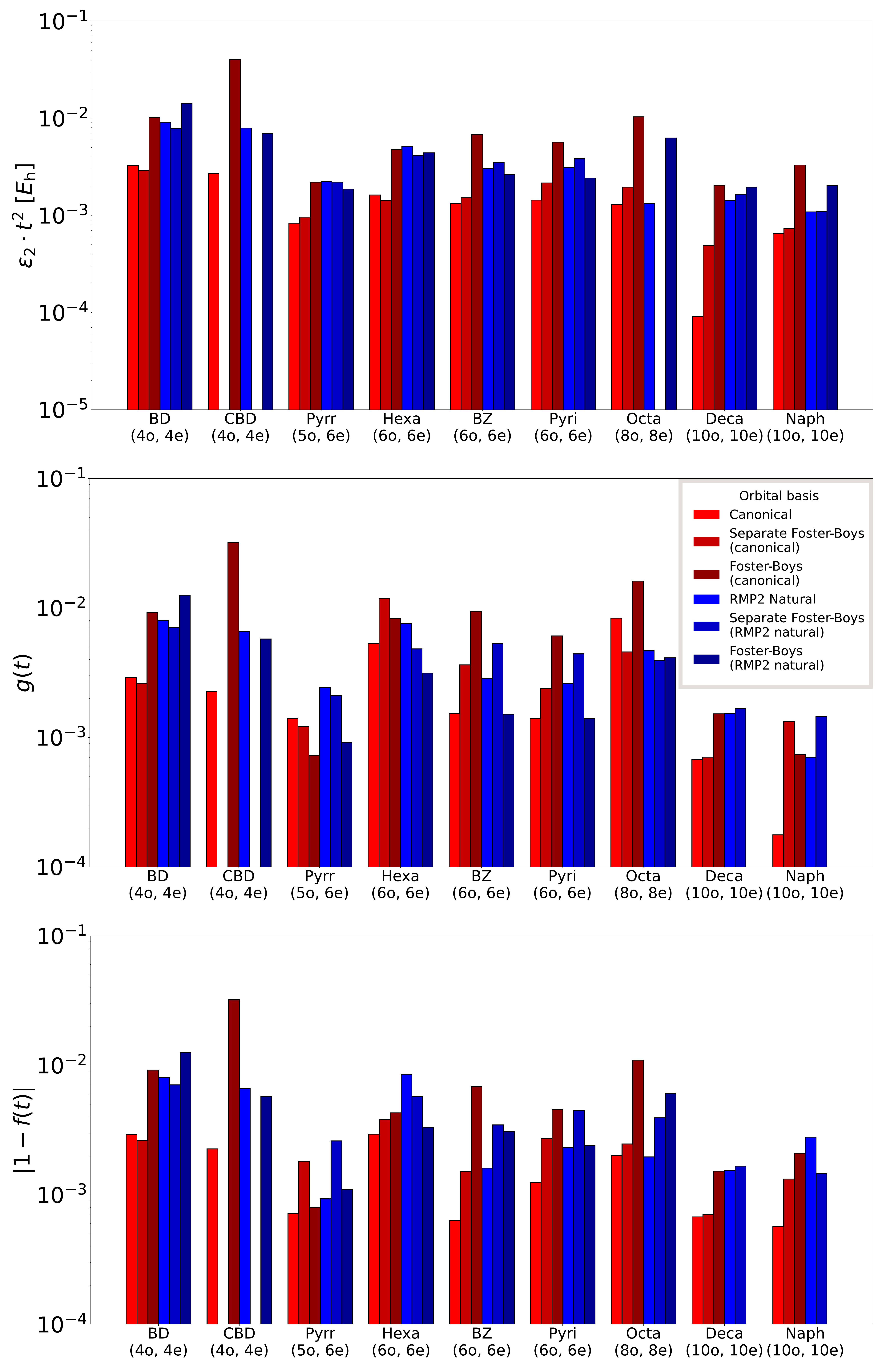}
    \caption{Estimates of Trotter error
    for the investigated $\pi$-systems.
    The fermionic Hamiltonian representation and magnitude
    ordering of the Trotter series were employed.
    Top: perturbation theory energy error estimate
    $\epsilon_2 t^2$.
    Middle: ACF offset $g(t)$.
    Bottom: infidelity $|1-f(t)|$.}
    \label{fig:correlating_quantites_larger_pi_sys}
\end{figure}

We now study the correlation
of $\lvert \Delta E_0 \rvert$ and various
estimates of Trotter error.
For a direct comparison, we refer to Fig. \ref{fig:med_orb_vs_dE}
for data on $\lvert \Delta E_0 \rvert$, Fig. \ref{fig:correlating_quantites_larger_pi_sys} for
$\epsilon_2 t^2$, $f(t)$ and $g(t)$, and to the SI for a plot of $\alpha t^2/2$.
We measured correlations by the Pearson correlation
coefficient $R^2$ between Trotter error $\lvert \Delta E_0 \rvert$ and
target error estimates, and considered all data points in the data
set of $\pi$-systems displayed in Fig. \ref{fig:med_orb_vs_dE},
i.e. $\pi$-systems with up to six spatial orbitals.
For $\epsilon_2 t^2$, $f(t)$, and $g(t)$ we found
significant correlation, with $R^2$ values of
0.656, 0.557, and 0.818, respectively.
However, poor correlation was observed with $\alpha t^2/2$ ($R^2=0.078$).
Accordingly, the perturbative error estimate $\epsilon_2 t^2$ as well as
the error measures in the time-evolved wave function $f(t)$
and $g(t)$ are suitable estimates of Trotter error $\lvert \Delta E_0 \rvert$.
We performed calculations of these three error estimates
on $\pi$-systems with up to 10 spatial orbitals.
Our results are shown in Fig. \ref{fig:correlating_quantites_larger_pi_sys}.

For the majority of cases, the employed energy error
estimates agree on canonical orbital bases resulting in lower Trotter errors
compared to localised bases.
The estimate $\epsilon_2 t^2$ predicts that the Trotter
error can vary by up to an order of magnitude
between different orbital bases
for $\pi$-systems with up to six spatial orbitals.
This is consistent with results of Trotter error
$\lvert \Delta E_0 \rvert$ in Fig. \ref{fig:med_orb_vs_dE}.
The estimate $\epsilon_2 t^2$ predicts that Trotter
error incurred in different orbital bases can vary
by up to more than one orders of magnitude
if systems with up to ten spatial orbitals are considered.
We note, however, that the correlation between $\lvert \Delta E_0 \rvert$ and $\epsilon_2 t^2$
was decreased for a qubit Hamiltonian representation. 

\subsubsection{Accounting for discrepancies in the ordering of terms in the Trotter series}\label{subsubsec:order_discrepancy}

\begin{figure}[H]
    \centering
    \includegraphics[width=0.9\linewidth]{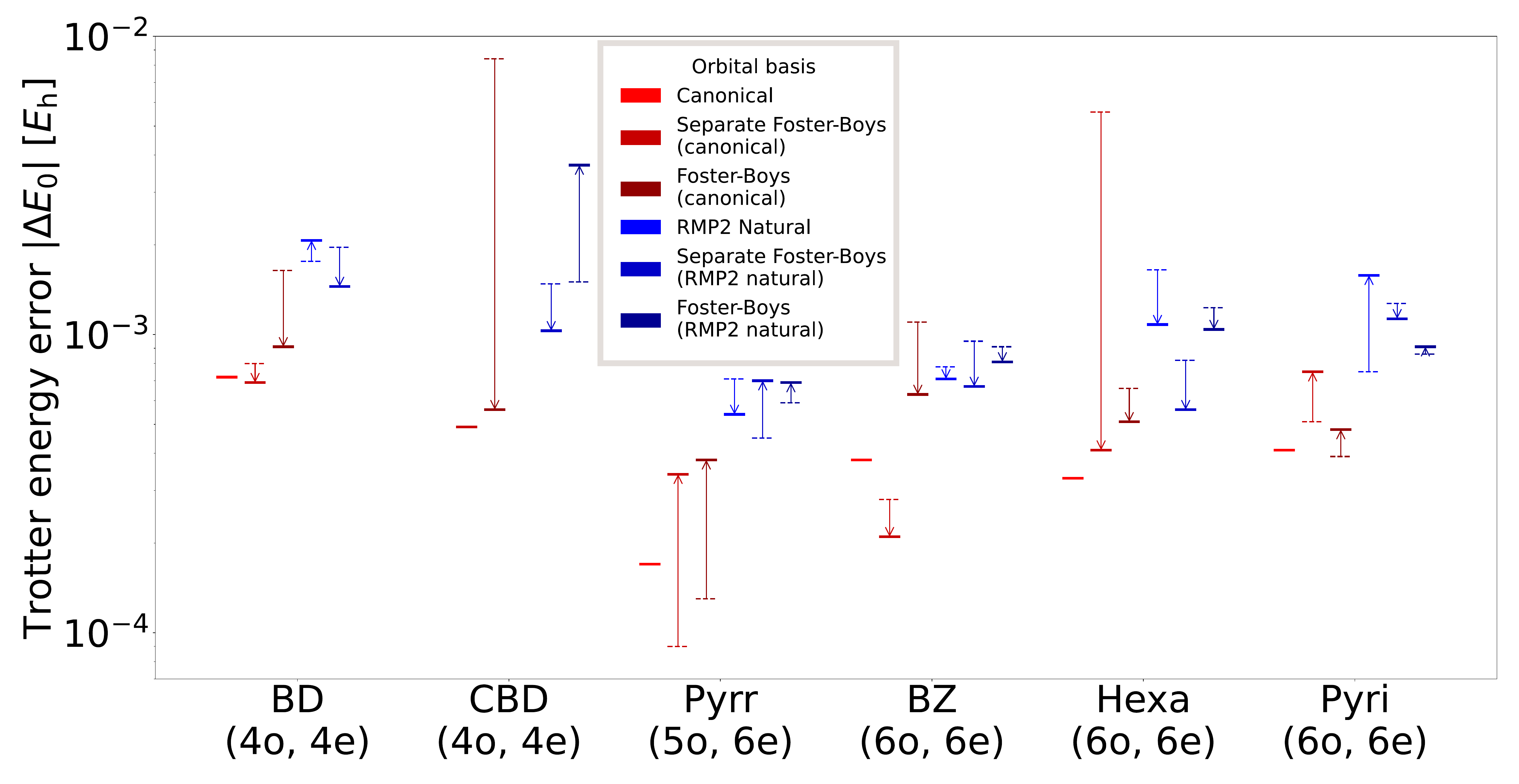}
    \caption{Trotter errors $\lvert \Delta E_0 \rvert$ obtained to precision
    $\epsilon=10^{-5}$ $E_{\mathrm{h}}$ (see Eqs. \eqref{dE_to_precision_epsilon}-\eqref{E_0_Trotter_via_IFT})
    for active spaces of various $\pi$-systems in different orbital bases.
    Thin, dashed bars: a (basis-dependent) magnitude ordering of the Trotter series
    was employed (data previously shown in Fig. \ref{fig:med_orb_vs_dE}).
    Thick bars: the terms in the Trotter series were reordered to match the
    magnitude ordering of the canonical orbital basis.}
    \label{fig:dE_trotter_reo}
\end{figure}

\begin{figure}[H]
    \centering
    \includegraphics[width=0.9\linewidth]{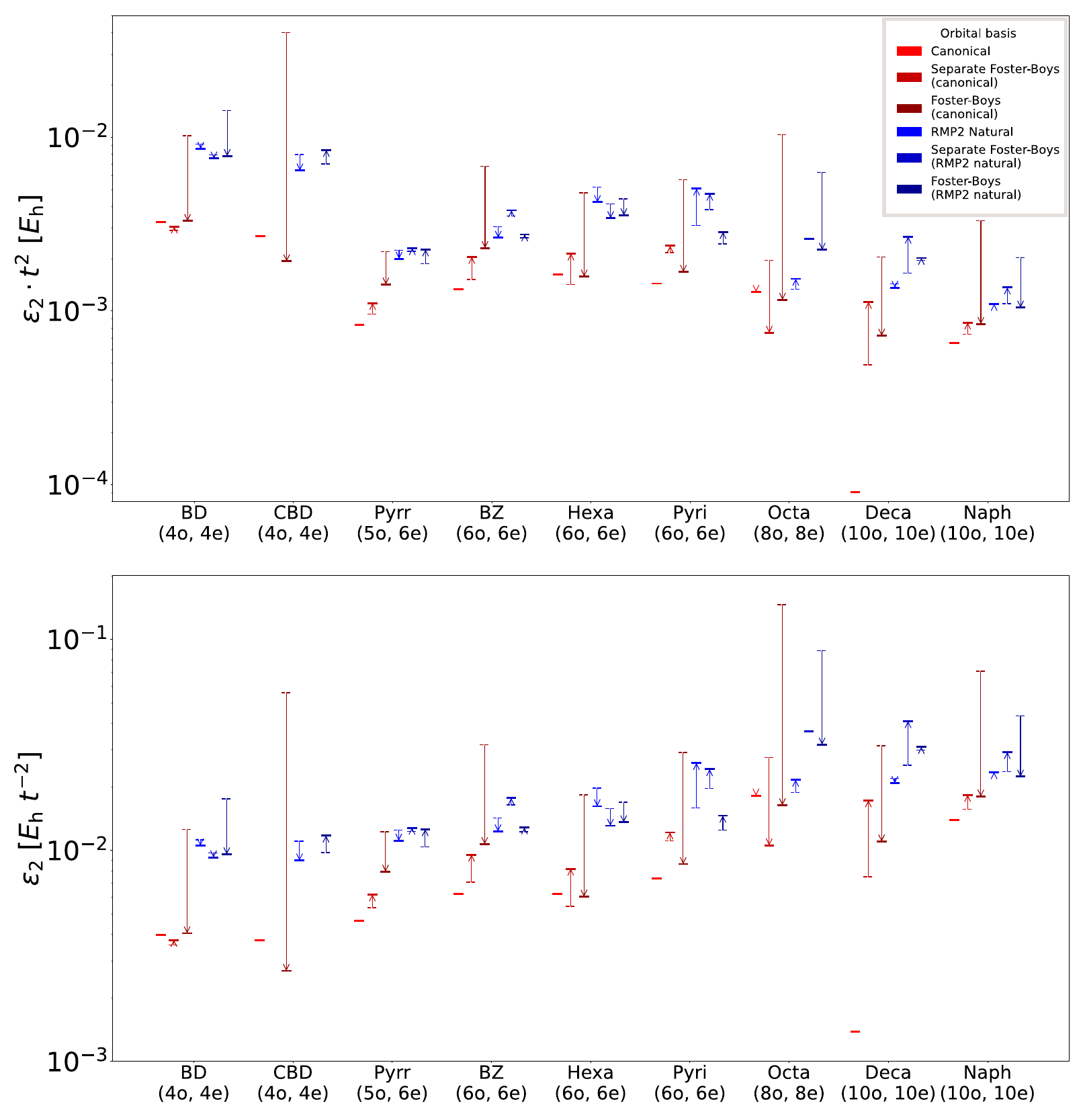}
    \caption{Trotter energy error estimates $\epsilon_2 t^2$  (top)
    and time-independent Trotter energy error estimates $\epsilon_2$ (bottom)
    for active spaces of various $\pi$-systems in different orbital bases.
    Thin, dashed bars: a magnitude ordering of the Trotter series
    was employed (data previously shown in Fig. \ref{fig:correlating_quantites_larger_pi_sys}).
    Thick bars: the terms in the Trotter series were reordered to match the
    magnitude ordering of the canonical orbital basis.}
    \label{fig:V2_dE_PT_reo}
\end{figure}

We now assess the effect of differences in the Trotter series-ordering
between the different orbital bases.
While our previous choice of a magnitude ordering of the Trotter
series was sensible, given that it typically incurs low Trotter errors \cite{TrotterOrderingImpact},
it resulted in a basis-dependent ordering of the terms $\op{H}_k$,
since the weights $\lvert h_k \rvert$ depend on the orbital basis.
However, to isolate the effect of the orbital basis on Trotter error,
a consistent Trotter series-ordering should be employed.
We previously observed that a canonical orbital basis often incurred the lowest Trotter error.
Hence, we performed calculations where the order of operators $\op{F}_k$ in the Trotter series
matched that of the magnitude ordering of the canonical basis (see SI
for additional information on this reordering);
results are shown in Figs. \ref{fig:dE_trotter_reo} and \ref{fig:V2_dE_PT_reo}.

The effect of changing the Trotter series-ordering
to the magnitude ordering of the canonical basis,
both on the Trotter error $\lvert \Delta E_0 \rvert$ as well as
the estimate $\epsilon_2$, can be notable.
Up to an order of magnitude differences can be observed,
and the largest differences are observed for a localised basis.
For a fixed molecular system, both $E_0$, and $\epsilon_2$ now
differ notably less between the different orbital bases
upon consistent ordering of the operators $\op{F}_k$.
(Localised) canonical orbital bases now display consistently
lower Trotter error compared to (localised) RMP2 natural orbital bases.
The previous observation that a localised basis
can substantially increase Trotter error can hence
largely be attributed to differences
in the Trotter series-ordering.
Still, even with a basis-independent ordering
of the Trotter series, an orbital basis which consistently
incurs the lowest Trotter errors cannot be identified.
However, the small variance of Trotter error in the different orbital
bases suggests that the effect of the orbital basis
on Trotter error can be much smaller compared
to prior results \cite{BabbushOrbInfluence}

We briefly comment on the growth of Trotter error with
the number of active spatial orbitals $N$:
Results for $\lvert \Delta E_0 \rvert$ and $\epsilon_2 t^2$ 
in Figs. \ref{fig:dE_trotter_reo} and \ref{fig:V2_dE_PT_reo} suggest
a decrease of Trotter error with system size.
Note, however, that both of these quantities depend on the
chosen Trotter step number $s$.
As listed in Table \ref{tab:system_oview}, we chose the time $t$
per Trotter step to be proportional to the inverse of
the ground state energy $E_0$.
It was also observed in Ref. \cite{TrotterOrderingImpact}
that choosing a Trotter step number $s$, such that
$\frac{1}{s} \propto t \propto \frac{1}{E_0}$
can lead to Trotter errors (measured in Ref. \cite{TrotterOrderingImpact}
by a perturbative error estimate from Ref. \cite{poulin2015trotter}) that is
decreasing with $N$.
By contrast, a time-independent measure 
of Trotter error, such as $\epsilon_2$ (see Fig. \ref{fig:V2_dE_PT_reo}),
shows that Trotter error (for fixed $t$) grows with $N$.

\subsubsection{Correlation of descriptors with Trotter error}\label{subsubsec:results_med_pi_target_quant_correlation}

We summarise results 
on the correlation between estimates of Trotter error (or exact Trotter error)
and easily obtainable descriptors of Trotter error
in Table \ref{tab:correlation_of_estimators_with_target_quants}.

\begin{table}[H]
    \centering
    \begin{tabular}{c|c|c}\hline\hline
         Trotter error (estimate) & 1-norm $\lambda_F$ & \# of terms $\Gamma$\\
         \hline
         $\alpha^{*}$, Eq. \eqref{childs_bound_alpha_def} \cite{ChildsTrotter} & 0.836 & 0.652 \\
         $\lvert \Delta E_0 ^{*} \rvert$ \eqref{target_eval_error} & 0.094 & 0.048 \\
         $\epsilon_2^{*}$, Eq. \eqref{epsilon_2_def} \cite{mehendale2025estimating} & 0.004 & 0.017 \\
         $\epsilon_2^{*} \cdot  t^2$ & 0.187 & 0.136 \\
         $\epsilon_2^{\mathrm{all}}$ & 0.164 & 0.337 \\
         $\epsilon_2^{\mathrm{all}} \cdot t^2$ & 0.139 & 0.092 \\
         \hline
         $\lvert \Delta E_{0} ^{*, \mathrm{cb}} \rvert$ & 0.071 & 0.091 \\
         $\epsilon_2^{*, \mathrm{cb}}$  & 0.426 & 0.251 \\
         $\epsilon_2^{*, \mathrm{cb}} \cdot t^2$  & 0.158 & 0.184 \\
         $\epsilon_2^{\mathrm{all, cb}}$  & 0.374 & 0.296 \\
         $\epsilon_2^{\mathrm{all, cb}} \cdot t^2$ & 0.296 & 0.249 \\
         \hline\hline
    \end{tabular}
    \caption{Pearson correlation coefficient $R^2$ of 1-norm $\lambda_F$ and number of terms in the
    Hamiltonian $\Gamma$ with various quantities describing the Trotter error
    in the energy: time-independent prefactor of the upper bound on Trotter error $\alpha$,
    perturbation theory error estimate $\epsilon_2$ and exact
    Trotter energy error $\lvert \Delta E_0 \rvert$.
    The asterisk denotes that only $\pi$-systems
    with up to six spatial orbitals were considered.
    The superscript '${\mathrm{all}}$' denotes that calculations
    on all $\pi$-systems (up to ten spatial orbitals) were considered.
    The superscript '${\mathrm{cb}}$' indicates that the canonical basis
    magnitude ordering was employed instead of the basis-dependent
    magnitude ordering.}
    \label{tab:correlation_of_estimators_with_target_quants}
\end{table}

A notable correlation is observed between
$\alpha$, which is proportional to the upper bound on Trotter error,
and both the 1-norm $\lambda_F$
and the number of terms $\Gamma$ in the Hamiltonian,
as suggested by Eq. \eqref{childs_bound_alpha_def}.
However, no significant correlation of Trotter error
$\lvert \Delta E_0 \rvert$ or of the estimate $\epsilon_2$
with $\lambda_F$ is evident.
Even a comparison of results with the canonical basis
and separately localised basis, which preserves the single-reference
character of $\ket{\psi_0}$, does not show a consistent reduction of Trotter error with the 1-norm.
The results shown in Table \ref{tab:correlation_of_estimators_with_target_quants}
highlight that consistent ordering
of the Trotter series in terms of the operators $\op{H}_k$
improved the correlation with the descriptors.

We conclude that the optimisation of circuit
depth by a-priori selection of a low-Trotter-error
orbital basis appears unfeasible.
Descriptors such as the 1-norm and the number of non-negligible
terms in the Hamiltonian depend solely on the orbital basis choice.
They do not account for the notable dependence of
Trotter error on the Trotter series-ordering.
Consequently, it is challenging to find a descriptor
which (i) can be easily evaluated, 
(ii) be optimised
through orbital transformations, (iii) is correlated
with Trotter error, and (iv) is independent of the
Trotter series-ordering chosen.

\subsection{Error cancellation with \texorpdfstring{$\eta$}{η}-basis propagators
and \texorpdfstring{$\eta$}{η}-ordering propagators}\label{subsec:error_cancellation_numerical_display}

Next, we investigate Trotter errors of
$\eta$-basis propagators $U_{\mathrm{rot}}$, and $\eta$-ordering propagators
$U_{\mathrm{reo}}$ (see Eqs. \eqref{U_rot_def} and \eqref{U_reo_def}).
$\eta$-basis propagators only feature benefits if they achieve
either systematic error cancellation or an averaging effect of Trotter errors.
Error cancellation means that the Trotter error $\lvert \Delta E_0 \rvert$
associated with the propagator $U_{\mathrm{rot}}$ is smaller
than the Trotter error of all of its constituent propagators $U_{\mathrm{rot},n}$.
Error averaging will be seen if the Trotter error
associated with $U_{\mathrm{rot}}$ is below the average Trotter error
of the constituent propagators $U_{\mathrm{rot},n}$.
Achieving systematic error averaging would already be useful, since
it implies a lower Trotter error than that achieved in
trotterised Hamiltonian simulation in a
randomly selected fixed orbital basis.

In search of a simple error cancellation criterion,
we studied Trotter errors of $\eta$-basis propagators and initially set $\eta = 2$ (i.e., only two orbital bases were employed
in the construction of $U_{\mathrm{rot}}$).
As example systems, we considered (i) the HF molecule in a 6-31g basis set,
with a (2o, 2e) orbital active space chosen around the Fermi level of the canonical orbital basis
and (ii) molecular hydrogen in a minimal sto-3g basis set with a (2o, 2e) orbital active space,
which corresponds to the Full CI active space.
Since these active spaces contain only $N/2=2$ spatial orbitals,
a single Givens rotation angle $\theta_{pq} \equiv \theta$ is sufficient
to describe all possible orbital transformations
as described by Eqs. \eqref{u_r_via_givens_def} and \eqref{u_r_givens_spinorb_basis}.

\begin{figure}[H]
    \centering
    \includegraphics[width=0.85\linewidth]{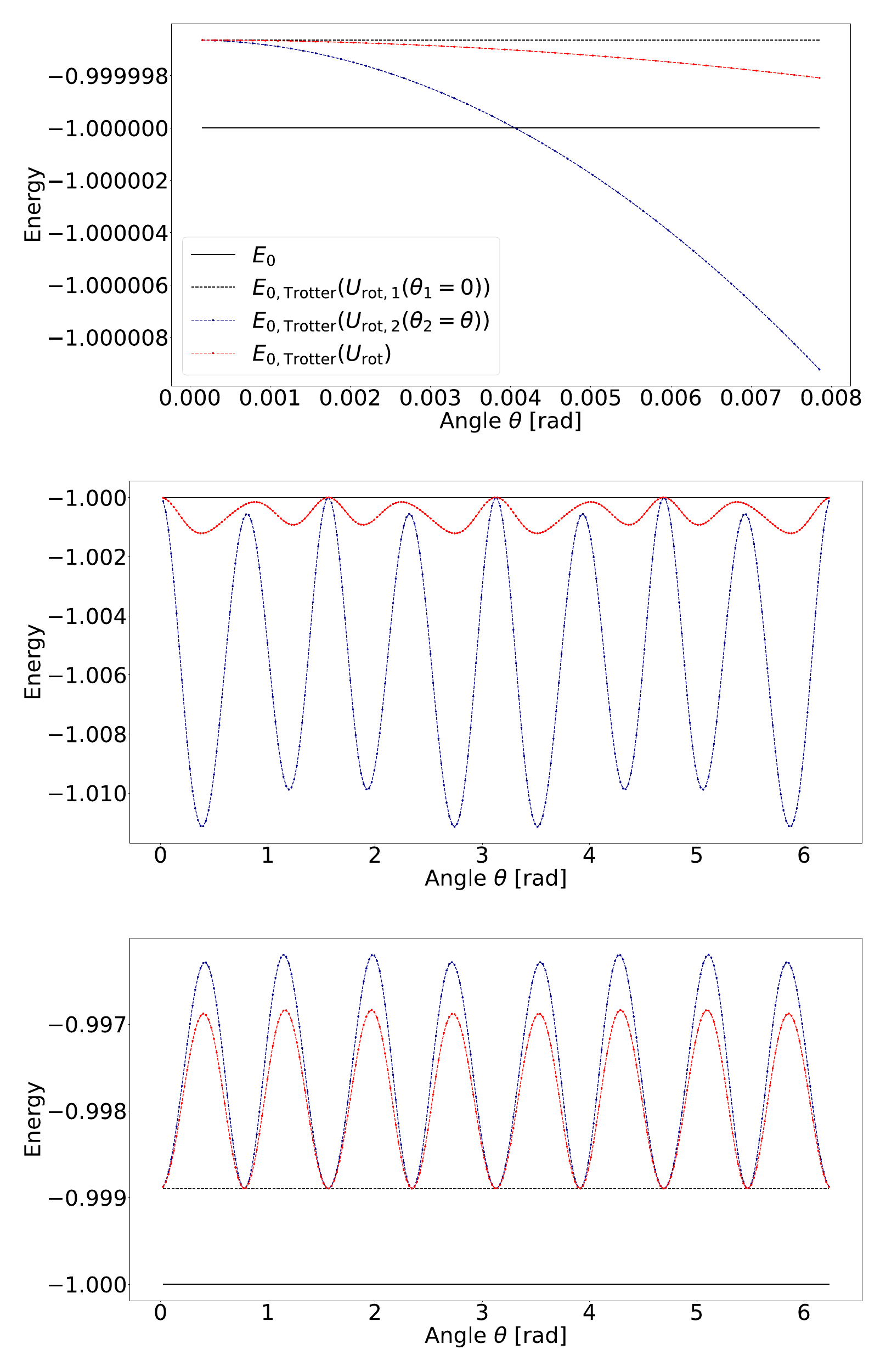}
    \caption{
    Ground state energies of the effective Hamiltonians
    of the $\eta$-basis propagator $U_{\mathrm{rot}}$
    and its constituent propagators $U_{\mathrm{rot}, 1}$ and $U_{\mathrm{rot}, 2}$.
    The angle $\theta$ denotes the orbital basis employed in
    the construction of $U_{\mathrm{rot}, 1}$ and $U_{\mathrm{rot}, 2}$.
    The canonical orbital basis serves as the reference for $\theta=0$.
    Normalised Hamiltonians were employed, such that the energy is unitless.
    The eigenvalue spectra of the Hamiltonians were shifted to the interval $[-1, 0]$,
    and the time evolution lengths was $t=0.95 \pi$.
    A fermionic Hamiltonian representation and the index-dependent
    Trotter series-ordering described in the SI were chosen.
    Top and middle: HF molecule at equilibrium bond length in a 6-31g basis
    with the active space chosen around the Fermi level.
    Top: small angle range for $\theta_2=\theta$. Middle: period $\theta \in [0, 2\pi)$ in the angle.
    Bottom: hydrogen molecule at equilibrium bond length in a sto-3g basis
    for a full period in the angle $\theta$.}
    \label{fig:interrot_error_cancel_test}
\end{figure}

For both examples ,we constructed propagators $U_{\mathrm{rot}}$
according to Eq. \eqref{U_rot_def} with $\eta=2$,
where $U_{\mathrm{rot}, 1}$ is constructed in the orbital basis labeled $\theta_1$
and $U_{\mathrm{rot}, 2}$ is constructed in the orbital basis labeled $\theta_2$.
We fixed the first orbital basis as $\theta_1 = 0$
and varied the second orbital basis $\theta_2$
on the interval $\theta_2 \in [0, 2 \pi)$.
We compare the ground state energies of the effective Hamiltonians
associated with the propagator $U_{\mathrm{rot}}$ and its constituents $U_{\mathrm{rot},1}$
and $U_{\mathrm{rot},2}$ in Fig. \ref{fig:interrot_error_cancel_test}.

As is apparent from Fig. \ref{fig:interrot_error_cancel_test},
the ground state energy $E_{0, \mathrm{Trotter}}$
associated with the propagator $U_{\mathrm{rot}}$ lies between that of its constituents
$U_{\mathrm{rot},1}$ and $U_{\mathrm{rot},2}$.
As such, error averaging is observed consistently.
A straightforward condition
required for error cancellation
which goes beyond an averaging effect can be observed:
the Trotter errors $\Delta E_0$ of the bases $\theta_1$
and $\theta_2$ are of opposite sign.

For the HF molecule in the interval of
$\theta_2 \in [0.006, 0.008]$, we can observe
that this leads to error cancellation.
For the hydrogen molecule, the error cancellation condition cannot be met,
since Trotter error has a positive sign for every possible orbital basis.

If combining orbital bases with opposite-sign Trotter error is a required condition
for error cancellation,
then Fig. \ref{fig:interrot_error_cancel_test}
suggests that error cancellation is unlikely to be
achieved by Hamiltonian simulation with randomised
orbital rotations between Trotter steps for the HF molecule.
This is because Trotter errors are
not symmetrically distributed around $\Delta E_0=0$
on the interval $\theta \in [0, 2 \pi)$ which contains
all orbital bases accessible by orbital transformations.
Additionally, for both the hydrogen molecule and the HF molecule,
the canonical orbital basis ($\theta=0$) incurs a small
Trotter error compared to a randomly selected basis
$\theta$.
Hence, for both molecules, we expect
Hamiltonian simulation in a fixed canonical basis to produce
smaller Trotter errors $\lvert \Delta E_0 \rvert$ compared
to $\eta$-basis propagators, which are constructed from $\eta$ randomly
chosen orbital bases.

\subsubsection{Distributions of Trotter error}\label{subsubsec:trotter_error_distributions}

\begin{figure}[H]
    \centering
    \includegraphics[width=0.9\linewidth]{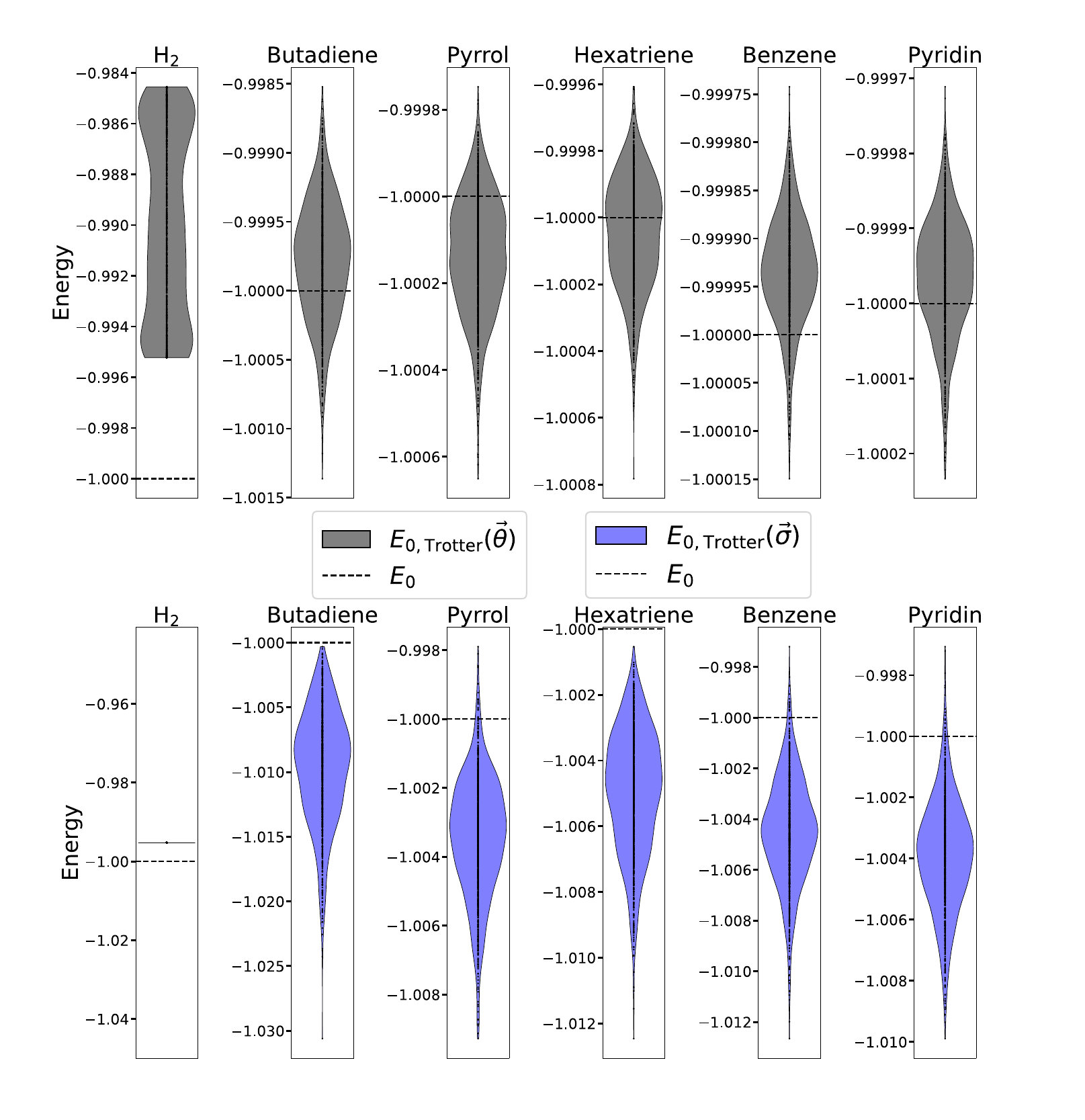} 
    \caption{Distribution of the ground state energy $E_{0, \mathrm{Trotter}}$
    of the effective Hamiltonian of $U_{\mathrm{Trotter}}$ in Eq. \eqref{H_eff_definition}
    obtained from sampling 1000 randomly selected orbital bases $\vec{\theta}$ (top)
    and 1000 randomly selected orderings $\vec{\sigma}$ of the Trotter series (bottom) for various molecules.
    Normalised, fermionic Hamiltonians were employed so that the energy is unitless.
    A time step of $t=0.95 \pi$ was employed.
    Eigenvalue spectra were shifted to the interval $[-1, 0]$.
    The exact ground state energy (dashed, horizontal black line) is shown as a reference.
    Black dots indicate individual data points.
    We include the hydrogen molecule in a sto-3g basis and the $\pi$-systems (with details on the active spaces provided in Table \ref{tab:system_oview}).
    In the construction of $\eta$-basis propagators, the index-based
    Trotter series-ordering described in the SI was applied.
    For $\eta$-ordering propagators, we chose the Foster-Boys localised canonical orbital basis
    for all molecules, except for the hydrogen molecule,
    for which the canonical basis was picked instead.}
    \label{fig:distributions_approx}
\end{figure}

Error cancellation might also be observed
for the $\eta(=2)$-ordering propagator $U_{\mathrm{reo}}$ in Eq. \eqref{U_reo_def},
if its constituents $U_{\mathrm{reo},1}$ and $U_{\mathrm{reo},2}$
yield Trotter errors of opposite sign.
However, for molecules with large active spaces $N$, we cannot determine
the sign of the Trotter error of an orbital basis $\vec{\theta}_n$
(or a Trotter series-ordering $\vec{\sigma}_n$), since $E_0$ is unknown.
We therefore require that a randomly selected set of orbital bases
$\{ \vec{\theta}_{n} \}_{n=1}^{\eta}$ (or orderings of the Trotter
series $\{ \vec{\sigma}_{n} \}_{n=1}^{\eta}$)
contains a similar number of orbital
bases (or orderings of the Trotter series) with positive
and negative Trotter error.
This necessitates that the distribution of Trotter error
in the different orbital bases (or the different orderings of the Trotter series)
itself is symmetrical around $\Delta E_0 = 0$.
To assess whether error cancellation can be achieved with randomly constructed $\eta$-basis propagators $U_{\mathrm{rot}}$
and $\eta$-ordering propagators $U_{\mathrm{reo}}$,
we therefore investigated the distributions of Trotter error in the orbital
basis $\vec{\theta}$ and the ordering of the Trotter series $\vec{\sigma}$. 
Our results are shown in Fig. \ref{fig:distributions_approx}.

As can be seen in Fig. \ref{fig:distributions_approx},
the Trotter error for the hydrogen molecule is unaffected
by the ordering of the Trotter series.
The distributions of $E_{0, \mathrm{Trotter}}$ are not symmetric
around $E_0$. Hence, the distributions of Trotter error $\Delta E_0$
are not symmetric around zero.
Among the examined distributions,
the distribution of Trotter error in the orbital basis with a fermionic
Hamiltonian representation is most symmetrical.
Surprisingly, negatively signed Trotter errors rarely occured,
independent of basis choice and ordering of the Trotter series
with a qubit Hamiltonian representation (see the SI).
As a result, we can generrally not expect trotterised propagators with random basis
transformations to reduce the Trotter error.

\begin{figure}[H]
    \centering
    \includegraphics[width=0.9\linewidth]{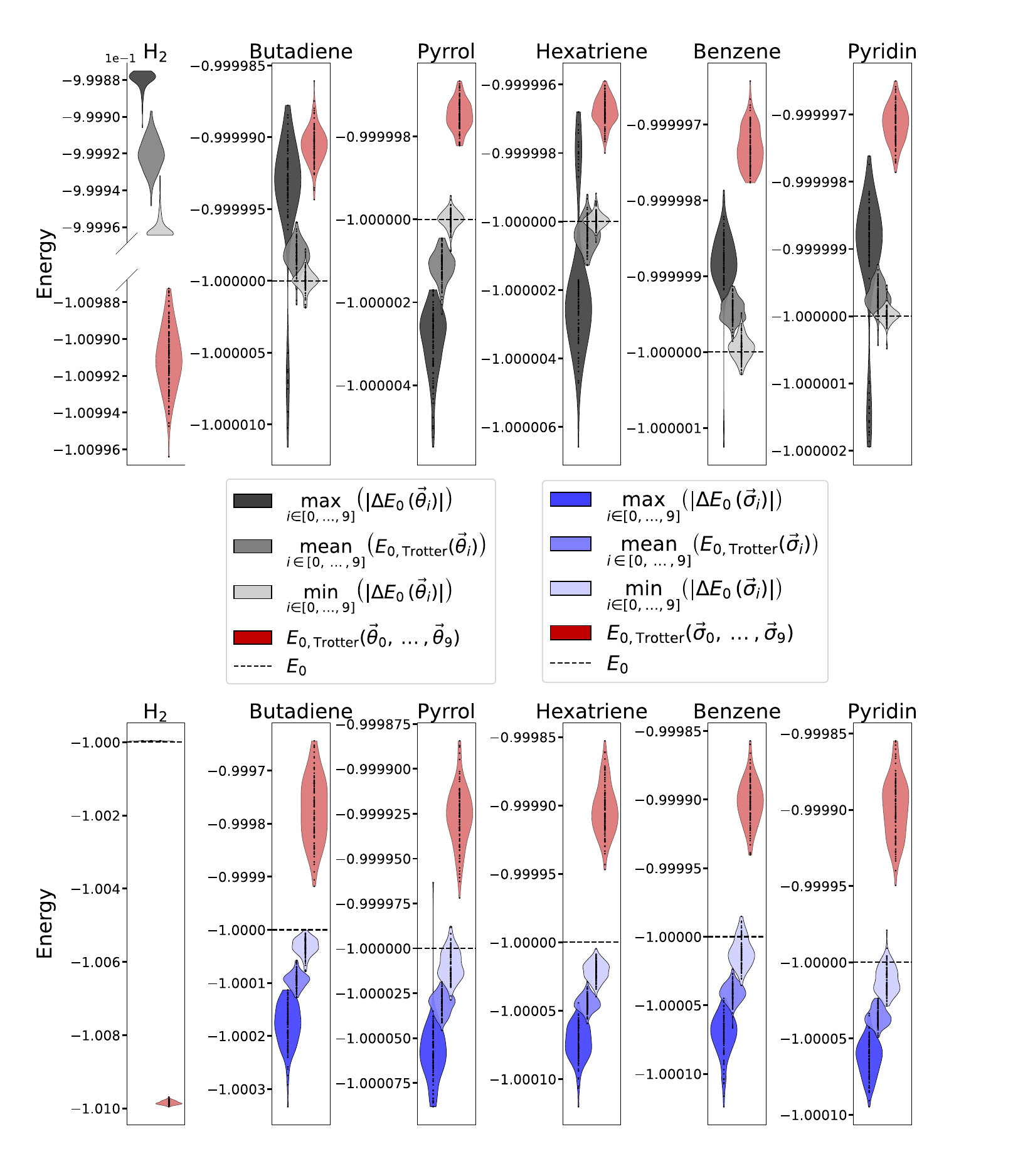}
    \caption{
    Top: Distributions of the ground state energy
    of the effective Hamiltonians of
    100 randomly constructed propagators $U_{\mathrm{rot}}$ (red).
    Distributions of the ground state energy of the effective Hamiltonians
    of the constituent propagators $U_{\mathrm{rot},n}$ that are
    associated with the largest
    (dark grey), average (grey), and
    smallest absolute value (light grey) 
    Trotter error.
    Bottom: Analogous comparison between randomised series
    ordering propagators $U_{\mathrm{reo}}$ (red) and their
    constituents $U_{\mathrm{reo}, n}$ exhibiting largest 
    (dark blue), average (blue), and
    smallest absolute value (light blue)
    Trotter error.
    For the construction of $\eta$-basis propagators $U_{\mathrm{rot}}$,
    the index-dependent Trotter series-orderings described in the SI
    was applied.
    For the construction of $U_{\mathrm{reo}}$, the chosen fixed basis $\vec{\theta}$
    was the Foster-Boys localised canonical orbital basis for all molecules, 
    except for molecular hydrogen, for which the canonical orbital basis
    was picked instead.
    Normalised Hamiltonians were employed so that the energy is unitless.
    An evolution time span of $t=0.95 \pi$ was set.
    Eigenvalue spectra were shifted to the interval [-1,0].
    Black dots indicate individual data points.
    Data are shown for a fermionic Hamiltonian representation.}
    \label{fig:periodic_driving_performance}
\end{figure}

\subsubsection{Trotter error of \texorpdfstring{$\eta$}{η}-basis propagators and
\texorpdfstring{$\eta$}{η}-ordering-propagators with \texorpdfstring{$\eta>2$}{η>2}} \label{subsubsec:random_propagators_practical_performance}

Finally, we investigate the Trotter error of $\eta$-basis propagators
constructed from randomly selected orbital basis
(and $\eta$-ordering propagators constructed from
randomly selected Trotter series-orderings)
with $\eta>2$.
For the molecules in Fig. \ref{fig:distributions_approx},
we constructed 100 distinct propagators $U_{\mathrm{rot}}$
(with $\eta=10$) from the constituent propagators $U_{\mathrm{rot},n}$.
The orbital basis $\vec{\theta}_n$ of each constituent propagator $U_{\mathrm{rot},n}$
was randomly selected.
Fig. \ref{fig:periodic_driving_performance} shows the
distribution of the ground state energies $E_{0, \mathrm{Trotter}}$
of the effective Hamiltonians of 100
$\eta$-basis propagators $U_{\mathrm{rot}}$.
$\eta$-ordering propagators $U_{\mathrm{reo}}$ were constructed
and assessed in an analogous fashion.

A shift towards positive-sign Trotter errors for the
$\eta$-basis propagators and $\eta$-ordering propagators is
observed for all systems (except for the dihydrogen).
This observation holds true for both the fermionic (Fig. \ref{fig:periodic_driving_performance}),
and for the qubit Hamiltonian representations (see the SI).
Additionally, the previous observation, namely
that error cancellation or averaging can be achieved by combination
of trotterised propagators, which incur oppositely signed
Trotter error, is inapplicable to $\eta$-basis propagators
and $\eta$-ordering propagators with $\eta$ larger than two.
This is apparent for the $\eta$-ordering propagators of
butadiene and hexatriene shown in Fig. \ref{fig:periodic_driving_performance}:
not a single propagator $U_{\mathrm{reo},n}$ in $U_{\mathrm{reo}}$ incurred a positively signed Trotter error.
A positively signed Trotter error is, however,
consistently observed for the propagators $U_{\mathrm{reo}}$.
For the fermionic Hamiltonian representation, we observe
that $\eta$-basis propagators $U_{\mathrm{rot}}$ incur a
Trotter error $\lvert \Delta E_0 \rvert$, which is
comparable to the largest Trotter error achieved by one
of the ten propagators $U_{\mathrm{rot},n}$ from which
$U_{\mathrm{rot}}$ is constructed.
A similar observation can be made for $\eta$-ordering propagators
$U_{\mathrm{reo}}$.
Consequently, neither error averaging, nor error cancellation
was observed for $\eta$-basis propagators or $\eta$-ordering propagators.
For the qubit Hamiltonian representation,
we consistently observe $U_{\mathrm{rot}}$
and $U_{\mathrm{reo}}$ to produce larger Trotter error
compared to all propagators $U_{\mathrm{rot},n}$
and $U_{\mathrm{reo},n}$ used for their construction.
Consequently, our results suggest that $\eta$-basis propagators
and $\eta$-ordering propagators lead to error magnification,
rather than error cancellation or averaging of the Trotter error.

\section{Conclusions}\label{sec:discussion}

We have assessed three strategies to reduce
ground-state-energy Trotter error by orbital basis transformations.

(i) We investigated whether the selection of a low Trotter error orbital basis
can be made before carrying out an actual quantum computation in order to reduce the resource cost of QPE.
To this end, we studied whether simple descriptors, which depend
solely on the employed orbital basis and which can be optimised
through orbital transformations, correlate with Trotter error.
Despite analytical expressions suggesting that these
descriptors should correlate with Trotter error,
poor correlation was observed in practice. This poor correlation we
attribute in parts
to their inability to account for the effect of
Trotter series ordering, known to significantly
affect Trotter error \cite{TrotterOrderingImpact}.
We found a reduced spread of ground-state Trotter error
for standard orbital bases of $\pi$-systems
compared to the atomic systems investigated in Ref. \cite{BabbushOrbInfluence}.
This suggests that the choice of a low Trotter error orbital
basis is of limited importance in practice.
Consequently, the reduction of the
gate count per Trotter step emerges as a more suitable avenue
to reduce Hamiltonian simulation cost for QPE.
It is already well-established that localised orbital bases
can reduce the gate count per Trotter step, since they
maximise the number of terms in the Hamiltonian that can be
neglected due to small coefficient weights \cite{LocalityQuantChem, OrbMin1norm}.

(ii) We observed that the Trotter error of the ground state energy is a continuous function of the transformed
orbital basis if an index-based Trotter series ordering is applied.
We proposed that this continuity might be exploitable
and allow for the determination of an orbital basis
in which the ground state energy is
free of Trotter error.
QPE with trotterised Hamiltonian simulation would
allow exact calculations of the ground state energy
in such an Trotter-error-free orbital basis.
However, if only an upper and lower bound
on $E_0$ are known, the determination of this
error-free basis can be challenging.

(iii) We investigated whether changing the orbital
basis between Trotter steps can reduce the Trotter error.
If we alternate the orbital bases between Trotter steps
where a negatively signed Trotter error
($\Delta E_0<0$) is combined with a positively signed Trotter error
($\Delta E_0>0$), error averaging naturally occurs,
and in some cases this leads to error cancellation.
However, for large active spaces,
we cannot determine whether an orbital basis has a positively or negatively signed Trotter
error prior to a quantum computation.
Additionally, results for trotterised propagators, which
employ up to ten different orbital bases
in their construction, suggest error magnification rather than
error cancellation or averaging for propagators
with a dynamical basis.
Hence, propagators which randomise the orbital basis are unlikely to
reduce Trotter error in practice.

If the derivation of tighter bounds on Trotter error by randomisation
of the orbital basis would be possible, trotterised Hamiltonian simulation
with a dynamical orbital basis could, however, still be useful if
a simulation with accuracy-guaranteed precision is desired.
A remaining possibility to explore orbital transformations
in the context of Hamiltonian simulation with product formulae is
the inclusion of orbital transformations in the qDRIFT protocol.
Whether this allows for the derivation of tighter
Trotter error bounds compared to a fixed orbital basis qDRIFT protocol
needs to be studied in future work.

We conclude that, while orbital transformations can be straightforwardly leveraged
to reduce the gate depth of a single Trotter step, consistent reductions in Trotter
error are very difficult to achieve by orbital transformation. Importantly, we could not find evidence that localised orbitals incur notably higher Trotter error compared to other choices of orbital basis sets.
Consequently, our results reconfirm the use of localised orbital basis for shallower QPE circuits.

\section*{Acknowledgements}\label{sec:acknowledgements}
We gratefully acknowledge generous financial support by the Swiss National Science Foundation (grant no. 200021\_219616) and by the Novo Nordisk Foundation (grant no. NNF20OC0059939 ‘Quantum for Life’).

\section*{Author contributions}
All authors contributed equally to the conceptualisation of this work.
M.K. and M.E. wrote the software and M.K. carried out the calculations.
All authors analysed the results and prepared the manuscript.
M.R. acquired funding, computing resources, and supervised the project.
M.R. is the corresponding author.

\section*{Disclosure statement}

The authors declare no conflict of interest.

\section*{Data availability statement}
The data that support the findings of this study will be made available
as a zenodo repository.

\printbibliography

\end{document}